\documentclass[preprint,12pt]{elsarticle}

\usepackage{enumitem}
\usepackage{color}
\usepackage{amsmath}
\usepackage{amssymb}
\usepackage{subcaption}
\usepackage{graphicx}
\usepackage{adjustbox}
\usepackage{graphics}
\usepackage{algorithm}
\usepackage{algpseudocode}
\usepackage{bbm}

\usepackage{adjustbox}

\usepackage{tabularx}

\usepackage{comment}

\usepackage{booktabs}
\usepackage{colortbl}
\usepackage{multirow}
\usepackage{fontawesome}

\usepackage{listings}

\usepackage{booktabs}
\usepackage[table]{xcolor}

\usepackage{orcidlink}

\journal{Computers in Biology and Medicine}

\begin{document}

\begin{frontmatter}



\title{Periodontal Bone Loss Analysis via Keypoint Detection With Heuristic Post-Processing}



\author[label1]{Ryan Banks\corref{cor1} \orcidlink{0009-0008-6504-7084}} 
\ead{rb01243@surrey.ac.uk}
\author[label1]{Vishal Thengane \orcidlink{0000-0002-3926-8405}} 
\author[label2]{María Eugenia Guerrero \orcidlink{0000-0001-5425-870X}}
\author[label3]{Nelly Maria García-Madueño \orcidlink{0000-0002-1333-8815}}
\author[label4]{Yunpeng Li \orcidlink{0000-0003-4798-541X}}
\author[label1]{Hongying Tang \orcidlink{0000-0003-2534-737X}}
\author[label5]{Akhilanand Chaurasia \orcidlink{0000-0002-8356-9512}} 

\cortext[cor1]{Corresponding Author}
\affiliation[label1]{organization={University of Surrey},
            addressline={Alan Turing Building}, 
            city={Guildford},
            postcode={GU2 7XH}, 
            state={Surrey},
            country={United Kingdom}}
\affiliation[label2]{organization={Universidad Nacional Mayor de San Marcos},
            addressline={Departamento Académico de Estomatologia Medico Quirurgico}, 
            city={Lima},
            postcode={WWW7+8H}, 
            state={Lima},
            country={Peru}}
\affiliation[label3]{organization={Universidad de San Martín de Porres},
            city={Lima},
            postcode={W2C2+6Q}, 
            state={Lima},
            country={Peru}}
\affiliation[label4]{organization={King's College London},
            addressline={Guy’s Tower, Guy’s Hospital}, 
            city={London},
            postcode={SE1 1UL}, 
            state={Greater London},
            country={United Kingdom}}
\affiliation[label5]{organization={King George’s Medical University},
            city={Lucknow},
            postcode={226003}, 
            state={Uttar Pradesh},
            country={India}}


\begin{abstract}
\noindent\textit{\textbf{Objectives:}} This study proposes a deep learning framework and annotation methodology for the automatic detection of periodontal bone loss landmarks, associated conditions, and staging.  

\noindent\textit{\textbf{Methods:}} $192$ periapical radiographs were collected and annotated with a stage agnostic methodology, labelling clinically relevant landmarks regardless of disease presence or extent. We propose a heuristic post-processing module that aligns predicted keypoints to tooth boundaries using an auxiliary instance segmentation model. An evaluation metric, Percentage of Relative Correct Keypoints ($PRCK$), is proposed to capture keypoint performance in dental imaging domains. Four donor pose estimation models were adapted with fine-tuning for our keypoint problem.

\noindent\textit{\textbf{Results:}} Post-processing improved fine-grained localisation, raising average $PRCK^{0.05}$ by $+0.028$, but reduced coarse performance for $PRCK^{0.25}$ by $-0.0523$ and $PRCK^{0.5}$ by $-0.0345$. Orientation estimation shows excellent performance for auxiliary segmentation when filtered with either stage 1 object detection model. Periodontal staging was detected sufficiently, with the best mesial and distal Dice scores of $0.508$ and $0.489$, while furcation involvement and widened periodontal ligament space tasks remained challenging due to scarce positive samples. Scalability is implied with similar validation and external set performance.

\noindent\textit{\textbf{Conclusion:}} The annotation methodology enables stage agnostic training with balanced representation across disease severities for some detection tasks. The $PRCK$ metric provides a domain-specific alternative to generic pose metrics, while the heuristic post-processing module consistently corrected implausible predictions with occasional catastrophic failures.

\noindent\textit{\textbf{Clinical significance:}} The proposed framework demonstrates the feasibility of clinically interpretable periodontal bone loss assessment, with potential to reduce diagnostic variability and clinician workload. 


\end{abstract}

\begin{graphicalabstract}
\textit{\textbf{Objectives:}} This study proposes a deep learning framework and annotation methodology for the automatic detection of periodontal bone loss landmarks, associated conditions, and staging.  

\noindent\textit{\textbf{Methods:}} $192$ periapical radiographs were collected and annotated with a stage agnostic methodology, labelling clinically relevant landmarks regardless of disease presence or extent. We propose a heuristic post-processing module that aligns predicted keypoints to tooth boundaries using an auxiliary instance segmentation model. An evaluation metric, Percentage of Relative Correct Keypoints ($PRCK$), is proposed to capture keypoint performance in dental imaging domains. Four donor pose estimation models were adapted with fine-tuning for our keypoint problem.

\noindent\textit{\textbf{Results:}} Post-processing improved fine-grained localisation, raising average $PRCK^{0.05}$ by $+0.028$, but reduced coarse performance for $PRCK^{0.25}$ by $-0.0523$ and $PRCK^{0.5}$ by $-0.0345$. Orientation estimation shows excellent performance for auxiliary segmentation when filtered with either stage 1 object detection model. Periodontal staging was detected sufficiently, with the best mesial and distal Dice scores of $0.508$ and $0.489$, while furcation involvement and widened periodontal ligament space tasks remained challenging due to scarce positive samples. Scalability is implied with similar validation and external set performance.

\noindent\textit{\textbf{Conclusion:}} The annotation methodology enables stage agnostic training with balanced representation across disease severities for some detection tasks. The $PRCK$ metric provides a domain-specific alternative to generic pose metrics, while the heuristic post-processing module consistently corrected implausible predictions with occasional catastrophic failures.

\noindent\textit{\textbf{Clinical significance:}} The proposed framework demonstrates the feasibility of clinically interpretable periodontal bone loss assessment, with potential to reduce diagnostic variability and clinician workload.

\end{graphicalabstract}

\begin{highlights}
\item We propose a keypoint annotation methodology for periodontal disease detection
\item It is stage agnostic and increases class counts, with staging done during inference
\item We propose a dental keypoint evaluation metric, Percentage of Relative Keypoints
\item We propose a heuristic post-processing method, to improve detections
\item Automating periodontal staging can reduce diagnostic time and improve accuracy
\end{highlights}


\end{frontmatter}



\section{Introduction}

Periodontal disease is an inflammatory condition that affects the gingiva and alveolar bone, a symptom of which is periodontal bone loss, which is progressive resorption of the alveolar bone supporting the teeth. If left untreated, periodontal bone loss results in tooth mobility and eventual tooth loss, which has major impacts on oral function, quality of life, and healthcare costs \cite{Jayakumar2010}. Periodontal disease affects between $20\%$ and $50\%$ of adults depending on the population studied \cite{helmi_prevalence_2019} and recent epidemiological surveys estimate that severe periodontitis affects nearly $10\%$ of the global population, ranking it among the most prevalent non-communicable diseases worldwide \cite{Eke2018WHO, Kassebaum2017GBD}.

The diagnosis of periodontal disease and the assessment of periodontal health in clinical practice, primarily relies on physical probing and radiographic evaluation \cite{periodontal2014kwthar}. Periodontal probing provides site-specific pocket depth measurements but is invasive and prone to error. Additionally, due the variability in patient anatomies, staging of periodontal disease cannot be accurately determined from physical probing alone. Radiographic analysis is used to further assess the extent and severity of periodontal bone loss and is the primary method of determining periodontal disease stages. Periapical radiographs provide a localised view of entire tooth anatomies and are the radiographic modality of choice for the assessment of periodontal bone loss for a set of target teeth, while panoramic radiographs offer a global overview of bone anatomical structures but introduce additional artefacts and distortions.

Staging of periodontal disease from radiographs is typically performed by measuring the tooth-aligned distance between the cementoenamel junction (CEJ), bone level (BL), and root level (RL). These distances are used to calculate the proportion of bone loss (PBL) relative to root length, where a healthy bone level is considered to be up to $2mm$ below the CEJ depending on age and lifestyle. PBL calculations, in conjunction with other clinical and radiographic indicators such as furcation involvement, widened periodontal ligament space (PLS), alveolar bone resorption (ARR), bleeding on probing, and plaque retention, are used to diagnose and stage periodontal disease \cite{chapter2021gov,Caton2018Staging}. At present, these assessments are conducted manually, making them time-consuming and dependent on clinical expertise. Radiographic assessment of periodontal bone loss can therefore be refined into three core tasks: (i) localisation of per-tooth anatomical landmarks such as CEJ, BL, and RL; (ii) identification of disease-related conditions including furcation involvement, ARR, and PLS detachment; and (iii) tooth orientation estimation for accurate PBL calculation. If sufficiently integrated with deep learning methodologies, these tasks can enable automated, standardised, and scalable assessment of periodontal bone loss in radiographic imaging.

Advancements in machine learning have significantly contributed to the field of medical imaging and diagnostics \cite{john2023automatic,wang2022medclip,wu2023medical}, enabling automation and improving the accuracy of disease diagnosis and analysis of conditions \cite{ronneberger2015u,valanarasu2021medical,banks2024hfcbformer,myles2024leveraging,jaamour2023divide,patra2018sequential}. Object detection and keypoint detection models, in particular, have demonstrated considerable promise in medical applications, including the identification of tumours in radiographic images and the detection of anomalies in ultrasound scans \cite{su14031447}. The integration of these technologies into periodontal bone loss detection has the potential to improve diagnostics by providing faster, more consistent, and widely accessible assessments.

Within dental domains, CNN-based architectures such as U-Net \cite{ronneberger2015u} are widely adopted for segmentation of anatomical structures \cite{eman2021novel,geetha2022panoramic} and pathology detection \cite{zheng2021improved,wei2021mau}. Attention-based models \cite{vaswani2017attention} have also been investigated in dental imaging \cite{dujic2023automatized}, though their reliance on large datasets remains a limitation. Recent studies have explored the automatic detection of periodontal disease using deep learning and signal processing methodologies \cite{xu2020dental,kubilay2022periodontal}, with some approaches applying keypoint detection for periodontal assessment \cite{vollmer2023periodontitis,liu2022pilot,jiang2022a,tsoromokos2022estimation,danks2021automating,cha2021determining}. These typically focus on detecting CEJ, BL, and RL landmarks to estimate PBL, while others have used segmentation to detect alveolar bone defects \cite{logiraj2024detsegdiff,chang2020deep}, object detection to directly detect periodontal disease \cite{thanathornwong2020automatic} or classify horizontal and vertical bone loss \cite{chang2022application}. However, existing methodologies remain limited: (i) keypoint-based works do not jointly incorporate all clinically relevant defect detection tasks, (ii) segmentation approaches do not consistently capture subtle conditions such as PLS detachment or ARR, (iii) current detection methodologies often produce anatomically implausible landmark predictions, and (iv) interpretability remains insufficient for clinical adoption \cite{SystematicReviewAPPRAISE2024}. Thus, there is a gap in research that unifys landmark localisation, disease-condition detection, and automated tooth orientation estimation for comprehensive periodontal assessment.

In this study, we introduce a deep learning framework that formulates periodontal bone loss assessment as a joint object detection, keypoint detection, and instance segmentation problem. We collect and annotate a dataset of $192$ intraoral periapical radiographs using a stage agnostic protocol that captures clinically relevant landmarks independently of disease severity or prevalence. This substantially increases class counts compared to direct detection methods, alleviating the data imbalance that often constrains periodontal imaging studies. Our proposed protocol also annotates additional periodontal disease related conditions such as deteched PLS, ARR and furcation involvement. The detection of these with additional tooth instance segmentation, providing tooth orientation estimation, enables the automatic calculation of PBL, detection of periodontal disease related conditions, and support standardised staging consistent with clinical guidelines.

We adapt four pose estimation architectures, Deep Pose \cite{toshev2014deeppose}, HRNet \cite{sun2019hrnet}, RTMPose \cite{jiang2023rtmpose}, and YOLOv8 \cite{jocher2023yolov8}, for landmark localisation and disease-related condition object detection. We also use YOLOv8-Seg fine-tuned on an auxiliary dataset for tooth instance segmentation. Using model detections, we propose a heuristic post-processing pipeline that aligns keypoints to relevant tooth boundaries, correcting anatomically implausible outputs that are common in existing approaches.

Finally, we propose an evaluation measure, Percentage of Relative Correct Keypoints ($PRCK$), which normalises localisation performance by the average tooth size within an image. Unlike generic pose estimation metrics, $PRCK$ ensures fair evaluation across teeth of varying morphology, making it more suitable for dental imaging tasks. Our results demonstrate that the framework can reliably detect key anatomical landmarks and conditions while improving localisation accuracy through post-processing, laying the foundation for automated, standardised, and clinically interpretable staging of periodontal disease.

\section{Methodology}
\label{sec:model_training}

\subsection{Dataset Annotation}
\label{sec:dataset_annotations}

The dataset consists of $192$ periapical radiographs, collected from $192$ patients representing a range of demographics and varying extents of periodontal health. A total of $582$ teeth are included in the dataset, comprising $386$ single-root teeth, $160$ double-root teeth, and $34$ triple-root teeth. From the $192$ periapical radiographs, $35$ images have healthy bone, $87$ have mild bone loss, $58$ have moderate bone loss and $12$ have severe bone loss.

The collection of radiographs was conducted by one periodontist (NG) and one radiologist (MG), both with at least $10$ years of clinical experience. Evaluation of periodontal disease extent was performed independently, without a time limit, and the evaluation periods were adjusted based on the availability of the observers. The annotators were tasked with classifying the area exhibiting the greatest bone loss into four categories: no bone loss, mild bone loss, moderate bone loss, and severe bone loss \cite{tonetti2018staging}. A consensus was reached between the two observers to assign the final category for each image.

\begin{figure*}[!htbp]
    \centering
    \begin{minipage}[b]{0.325\textwidth}
        \centering
        \includegraphics[width=\textwidth]{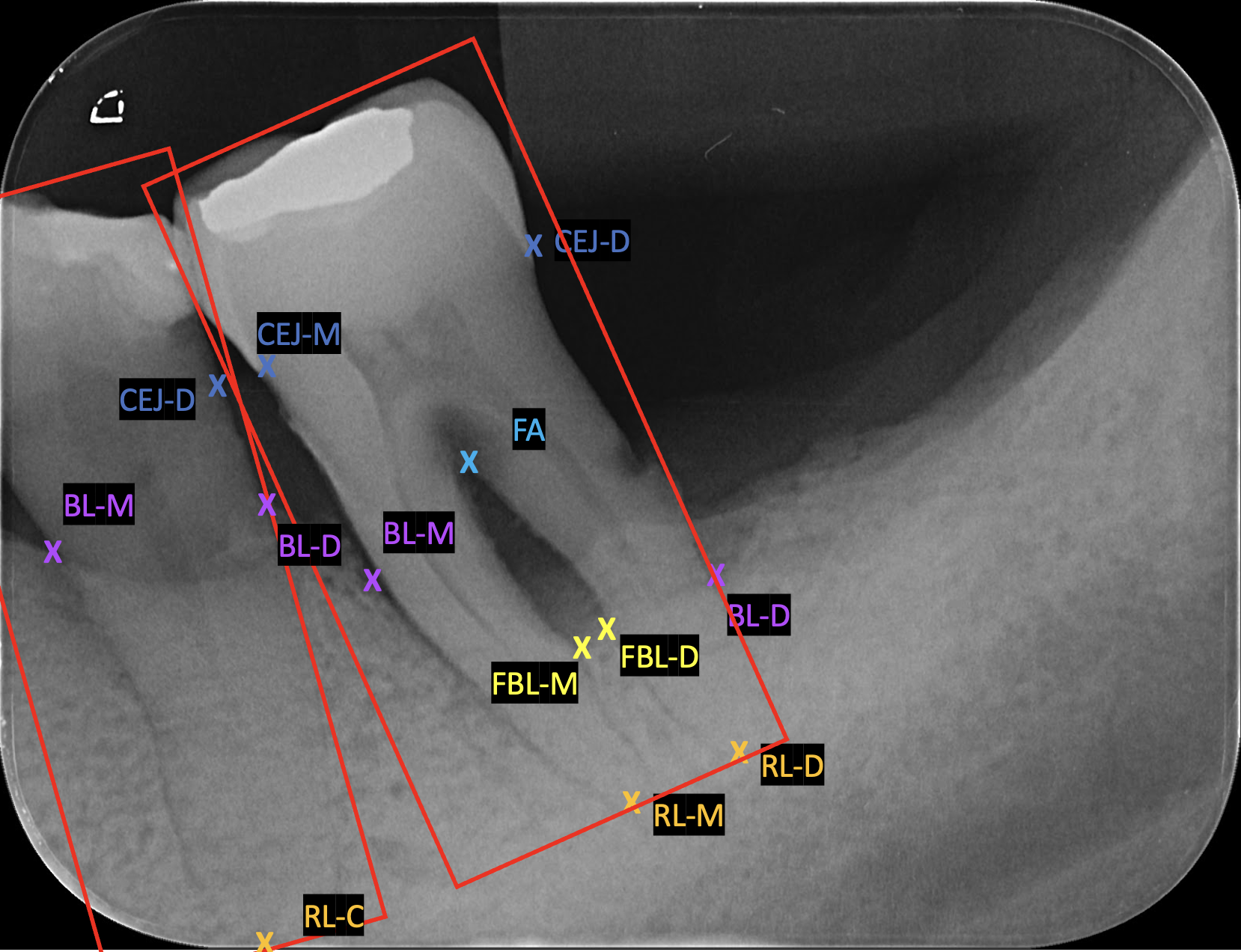} 
    \end{minipage}
    \hfill
    \begin{minipage}[b]{0.325\textwidth}
        \centering
        \includegraphics[width=\textwidth]{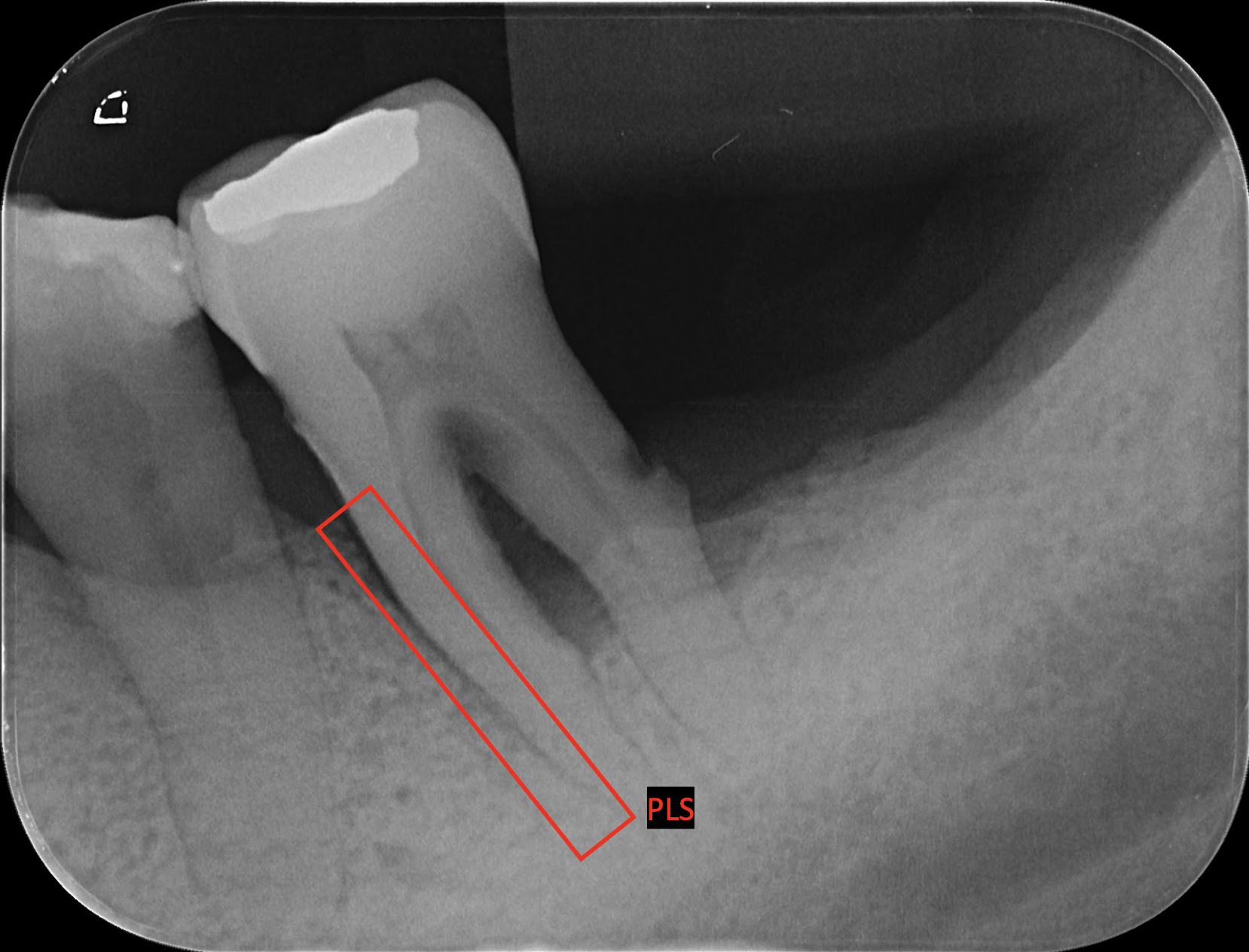} 
    \end{minipage}
    \hfill
    \begin{minipage}[b]{0.325\textwidth}
        \centering
        \includegraphics[width=\textwidth]{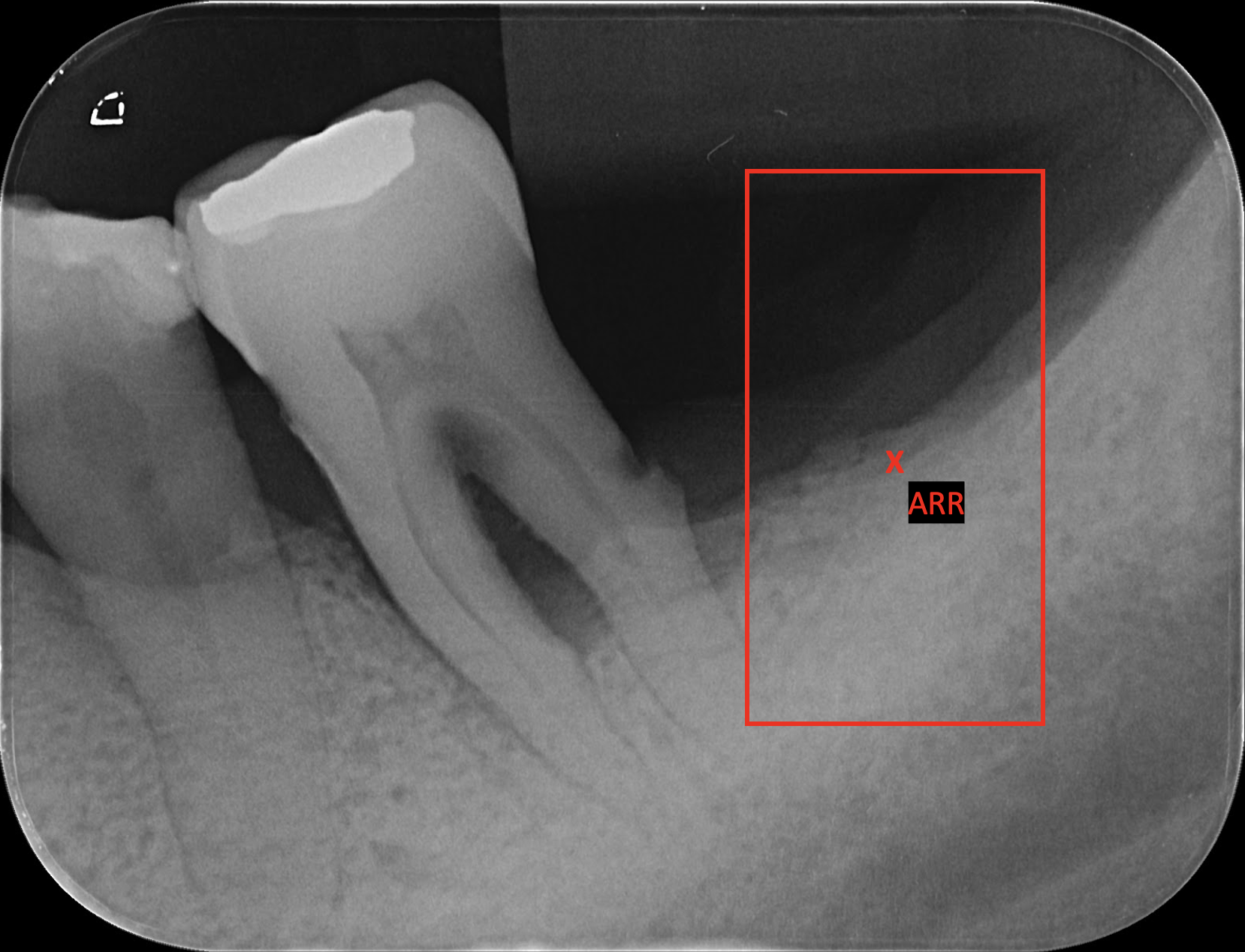} 
    \end{minipage}
    \caption{Three images containing example annotations of the collected keypoints and rotating bounding boxes.}
    \label{fig:annotation_disc}
\end{figure*}

We conducted comprehensive annotation of the collected radiographs, divided into four steps. These include annotating Bone Level Keypoints (BLK) for each tooth, identifying the Teeth Bounding Box (TBB) with tooth orientation, annotating ARR keypoints, and widened PLS bounding boxes. Figure \ref{fig:annotation_disc} provides visual examples of the annotations.

\textbf{Bone Level Keypoints (BLK)}\hfill\\
The first step involves labelling keypoints relating to the  Cementoenamel Junction (CEJ), current Bone Level (BL), and Root Level (RL) on both the mesial (-m) and distal (-d) sides of the teeth. For triple and single-root teeth, a central root level (RL-c) was also included, where single root teeth do not contain RL-m or RL-d keypoints. For multi-root teeth, additional keypoints were annotated to indicate furcation involvement. These included Furcation Apex (FA) and Furcation Bone Level mesial/distal (FBL-m, FBL-d), to indicate furcation involvement, and Furcation Bone Level Healthy (FBL-H) with FA, to indicate a healthy furcation area. FBL-h keypoints are not used by the model, but to identify indication of healthy furcation areas by the annotator, if furcation involvement keypoints are lost or missed in the annotation process. These annotations provide crucial information for assessing the extent of bone loss and periodontal disease, aligning the task with a computer vision problem.

\textbf{Alveolar Ridge Resorption (ARR)}\hfill\\
In the ARR annotation step, focus is placed on identifying the areas of Alveolar Ridge Resorption. This involves annotating the current bone level at locations where a tooth is missing and bone resorption has begun. Annotations for ARR are completed as a bounding box indicating the missing tooth area and a keypoint indicating the lowest point of ARR within the localised area of the missing tooth.

\textbf{Periodontal Ligament Space (PLS)}\hfill\\
In this step, areas where the periodontal ligament space had widened were annotated. PLS annotations in this category indicate ligament spaces that have widened from the tooth with a rotating bounding box, serving as indicators of compromised periodontal health rather than a healthy ligament space. 

\textbf{Teeth Bounding Box (TBB)}\hfill\\
The final step in the annotation process involved annotating rotating bounding boxes around each tooth. These serve as a reference box for the model detections of BLK locations and facilitates the calculation of bone loss percentages by identifying tooth orientation.

\subsubsection{Annotation Cleaning and Processing}
\label{sec:dataset_cleaning}

Once the annotations were completed, the dataset underwent a cleaning process to prepare it for training. Some annotations, such as ARR keypoints and bounding boxes for certain teeth, were either missing or misclassified. In addition, keypoints were not initially linked to their corresponding bounding boxes. To address these issues, keypoints were automatically matched to their bounding boxes by measuring the horizontal distance between each keypoint and the center of the each bounding box, choosing the shortest distance for each keypoint class. In cases where automatic matching failed, manual adjustments were made to correctly assign keypoints.

\begin{figure}[!htbp]
    \begin{subfigure}{0.47\columnwidth}
        \centering \includegraphics[width=\linewidth]{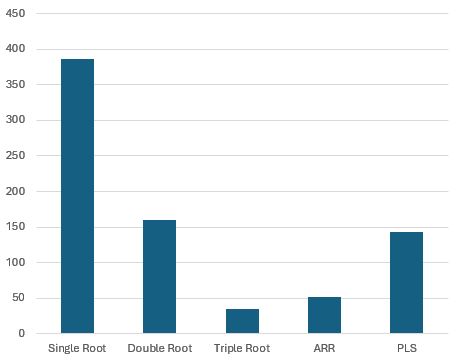}
        \caption{}
        \label{subfig:stat_a}
    \end{subfigure}
    \hfill
    \begin{subfigure}{0.47\columnwidth}
        \centering \includegraphics[width=\linewidth]{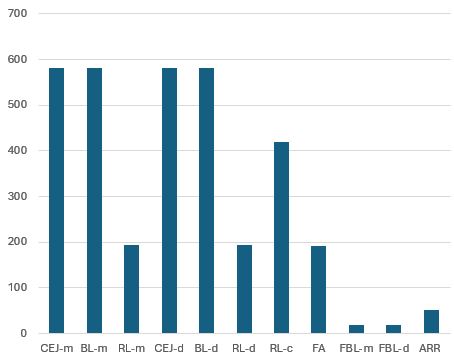}
        \caption{}
        \label{subfig:stat_b}
    \end{subfigure}

    \caption{(a): Bar plot showing the counts of bounding box class instances of our baseline dataset before processing and after cleaning. (b): Bar plot showing the counts of keypoint instances of our baseline dataset before processing and after cleaning.}
    \label{fig:statistics_plots}
\end{figure}

Figure \ref{fig:statistics_plots} presents the instance counts for each bounding box and keypoint class after cleaning. Figure \ref{subfig:stat_a} displays the counts for bounding boxes, and Figure \ref{subfig:stat_b} presents the keypoint counts. Some keypoint classes, such as Triple Root boxes, ARR boxes, FBL-m, and FBL-d keypoints, had low instance counts, which could pose challenges related to model overfitting.

Given the limitations of some current pose estimation models, which do not support rotation indices for bounding boxes, we removed the rotational component from the TBB and PLS bounding boxes. We instead determine rotation using predicted segmentation from an auxiliary tooth segmentation model later on. 

To ensure compatibility with the YOLOv8-Pose model, which requires all keypoints to be detected with visibility identifiers with assignment to a bounding box, we formatted the data into five bounding box classes: `Single Root', `Double Root', `Triple Root', `ARR', and `PLS'. Each keypoint for each box class was assigned a visibility value as follows: visibility 0 (not visible, not trained), visibility 1 (partially visible, trained), and visibility 2 (visible, trained). A summary of the visibility assignment for each keypoint depending on their attached bounding box class is shown in Table \ref{tab:dataset_classes}.

\begin{table}[!htbp]
    \centering
    \caption{Table displaying the visibility setting for each keypoint relating to each bounding box class. 0: Not Visible and Not Trained, 1: Partially Visible and Trained, 2: Visible and Trained.}
    \begin{adjustbox}{width=\columnwidth,center}
    \begin{tabular}{cccccccccccc}
         & CEJ-m  & BL-m  & RL-m  & CEJ-d & BL-d & RL-d & RL-c & FA & FBL-m & FBL-d & ARR\\
        Single Root & 2 & 2 & 0 & 2 & 2 & 0 & 2 & 0 & 0 & 0 & 0\\
        Double Root & 2 & 2 & 2 & 2 & 2 & 2 & 0 & 2 & 2 or 1 & 2 or 1 & 0\\
        Triple Root & 2 & 2 & 2 & 2 & 2 & 2 & 2 & 2 & 2 or 1 & 2 or 1 & 0\\
        ARR & 0 & 0 & 0 & 0 & 0 & 0 & 0 & 0 & 0 & 0 & 2\\
        PLS & 0 & 0 & 0 & 0 & 0 & 0 & 0 & 0 & 0 & 0 & 0\\
    \end{tabular}
    \end{adjustbox}
    \label{tab:dataset_classes}
\end{table}

CEJ and BL keypoints are present for all three of the tooth bounding boxes, with varying RL keypoints depending on the number of roots present in the tooth. For furcation involvement, healthy furcation areas were processed to include both FBL-m and FBL-d keypoints at the same location as the Furcation Apex (FA) keypoint, with visibility 1, ensuring the model could be trained on all instances, regardless of whether the furcation area was diseased. ARR keypoints are given their own ARR bounding box with ARR being the only keypoint trained. Additionally, no keypoints were used for the detection of PLS objects, as it is solely based on bounding box detection.

After completing the annotation and preprocessing steps, the final dataset consists of $192$ images, $578$ tooth bounding boxes, and a total of $3520$ keypoints. Prior to processing, the dataset included $19$ FBL-m and $19$ FBL-d keypoints indicating furcation involvement. However, after incorporating FBL keypoints for healthy furcation areas, the dataset now contains $191$ FBL-m and FBL-d keypoints.

\subsection{Heuristic Post-Processing}

Periodontal bone loss related keypoints such as the CEJ, BL, and RL keypoints, must exist along the edge of the related tooth. This is due to the staging guidelines for periodontal bone loss defining stages as alveolar bone loss compared to the length of the related tooth's root. This definition creates a heuristic rule on these keypoints that can be exploited. Therefore, if the pose detection model fails to predict the exact location of the keypoints along the tooth's edge, we can post-process the keypoints to realign to the edge of the tooth, improving keypoint predictions.

During the inference stage of the keypoint detection model, we propose a heuristic based post-processing module, that utilises prior knowledge and an independently pre-trained tooth segmentation model on an auxiliary panoramic tooth dataset. The segmentation model determines the outline of each tooth in the image and the module matches and adjusts the keypoint predictions to align along the edge of the relevant tooth, for a determined mesial or distal side of the tooth mask. 

The post-processing module is not used on any of the furcation related keypoints (FA, FBL-m, FBL-d) or the alveolar ridge resorption (ARR) keypoint. This is due to significant instances of overlapping roots in some double and triple root teeth, which makes the segmentation of the furcation area impossible, causing the post-processing to fail for these examples. Additionally, ARR keypoints indicate bone resorption in areas with missing teeth, so there are no nearby teeth to adjust the keypoint to.

\subsubsection{Segmentation Model Pre-training and Non-Maximum Merging}

\noindent To enable heuristic post-processing we need to find the edges of each independent tooth in the image. We fine-tuned YOLOv8-Seg on the open-source panoramic radiograph auxiliary dataset \cite{humans2023teeth}. We do not adjust the weights during keypoint training as the segmentation model is only used in the inference stage of or method. The panoramic auxiliary dataset consists of $598$ images annotated by 15 trainees from the Democratic Republic of Congo, with the images originating from patients of varying ages from Paraguay \cite{mello2021panoramic}. The panoramic radiograph dataset and our periapical dataset are from different sources, patients and radiography techniques. 

As a single trained segmentation model is to be used on our differing domain periapical dataset, which does not contain its own tooth segmentation labels, we primarily evaluated the performance qualitatively on our dataset and the cropped panoramic validation set. Additional training details and quantitative results on the auxiliary segmentation validation set are in \ref{app_seg_results}

\begin{figure}[htbp]
    \begin{subfigure}{0.40\columnwidth}
        \centering \includegraphics[width=0.70\linewidth]{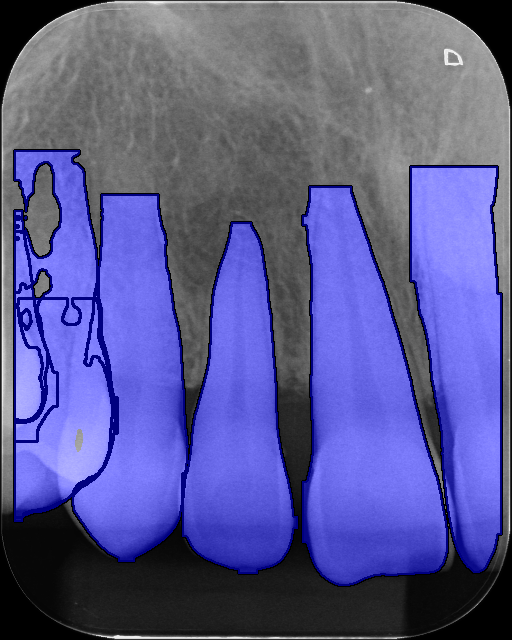}
        \caption{Before NMM}
        \label{subfig:1a}
    \end{subfigure}
    \hfill
    \begin{subfigure}{0.40\columnwidth}
        \centering \includegraphics[width=0.70\linewidth]{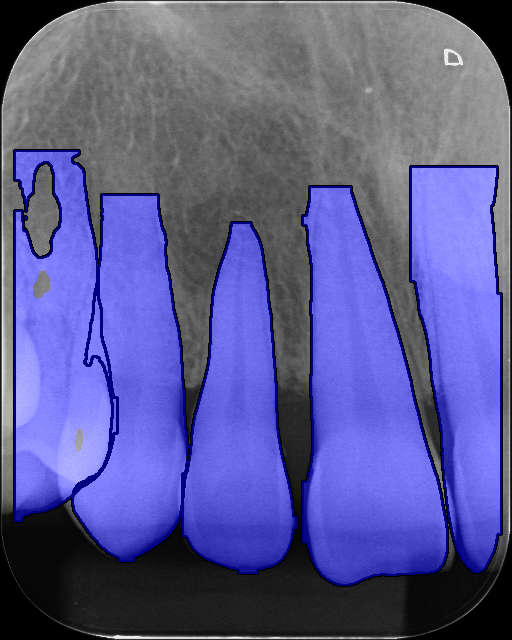}
        \caption{After NMM}
        \label{subfig:1b}
    \end{subfigure}

    \caption{Images containing predicted segmentation mask overlays, for Image 1, where (a) is before NMM and (b) is after NMM.}
    \label{fig:nmm}
\end{figure}

During inference with the post-processing module, we intentionally tuned prediction stage hyperparameters to over-predict segmentation masks. We did this by setting the Intersection over Union (IoU) threshold to $0.7$ and confidence threshold to $0.15$, which predicts a large number of overlapping masks of varying quality and completeness. We then use Non-Maximum Merging (NMM) \cite{kondrackis2024merging} with an IoU threshold of $0.1$. This is done to combine the large number of poor quality predicted masks, that is produced by the model when qualitatively evaluated on our different domain distribution dataset, into a cohesive and higher quality tooth masks, as seen in Figure \ref{fig:nmm}.

\noindent When segmenting the teeth in the image, we expect occasional false positive predictions from the instance segmentation model. This in turn, produces multiple redundant segmentation mask orientation angles, which are filtered out during post processing.

\subsubsection{Post-Processing Module Stages}

\begin{figure}[htb!]
\begin{center}
    \begin{subfigure}{0.32\columnwidth}
        \centering \includegraphics[width=1.00\linewidth]{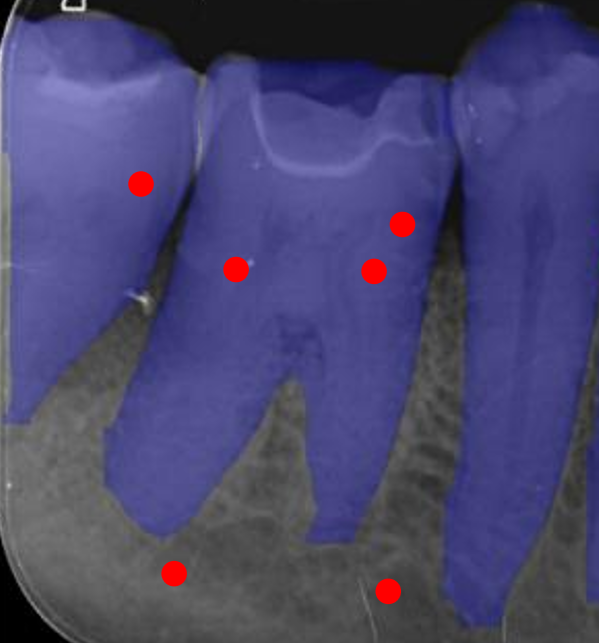}
        \caption{Predicted Keypoints/Masks}
        \label{subfig:postp1}
    \end{subfigure}
    \begin{subfigure}{0.32\columnwidth}
        \centering \includegraphics[width=1.00\linewidth]{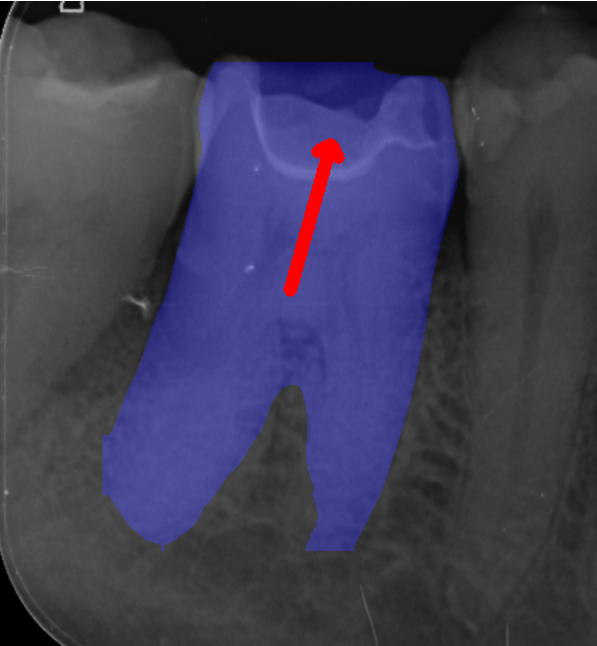}
        \caption{Tooth Orientation}
        \label{subfig:postp2}
    \end{subfigure}
    \begin{subfigure}{0.32\columnwidth}
        \centering \includegraphics[width=1.00\linewidth]{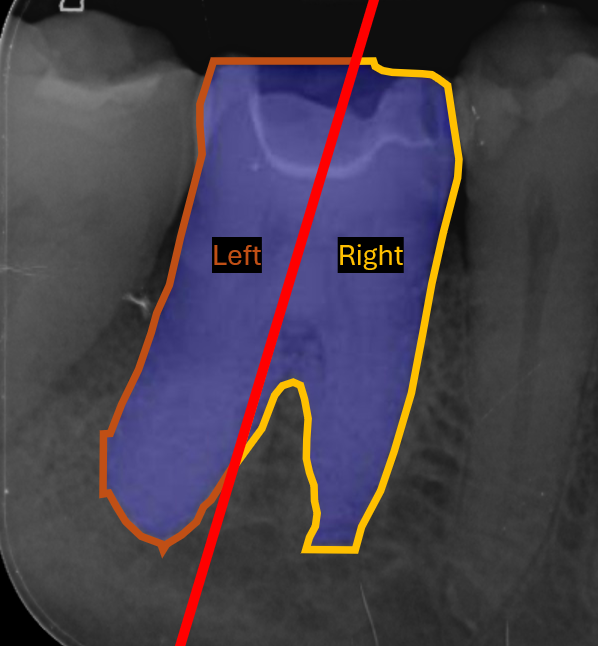}
        \caption{Mask Partitioning}
        \label{subfig:postp3}
    \end{subfigure}
    \newline
    \begin{subfigure}{0.32\columnwidth}
        \centering \includegraphics[width=1.00\linewidth]{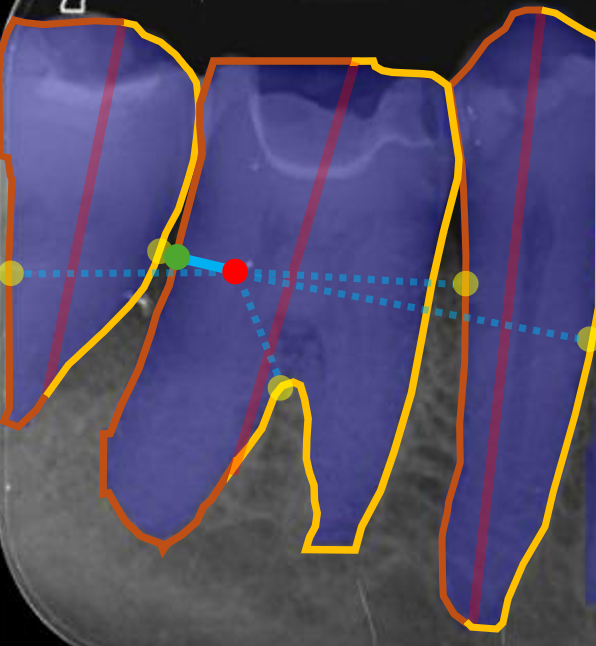}
        \caption{Edge Association}
        \label{subfig:postp4}
    \end{subfigure}
    \begin{subfigure}{0.32\columnwidth}
        \centering \includegraphics[width=1.00\linewidth]{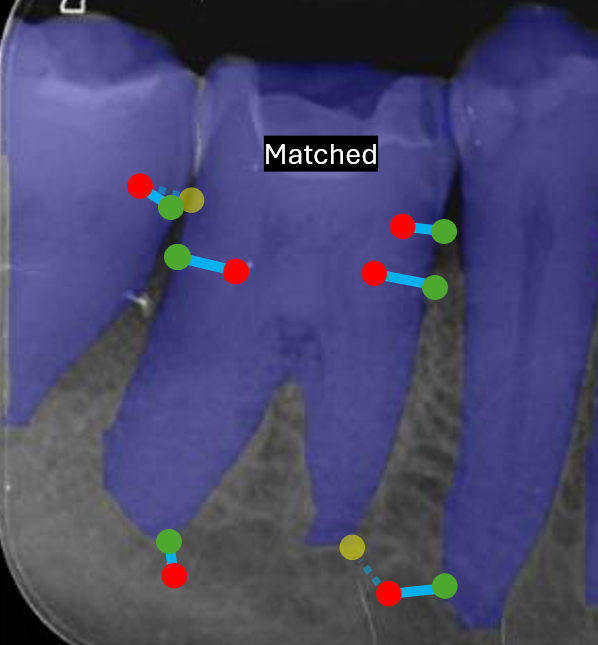}
        \caption{Mask-Keypoint Matching}
        \label{subfig:postp5}
    \end{subfigure}
    \begin{subfigure}{0.32\columnwidth}
        \centering \includegraphics[width=1.00\linewidth]{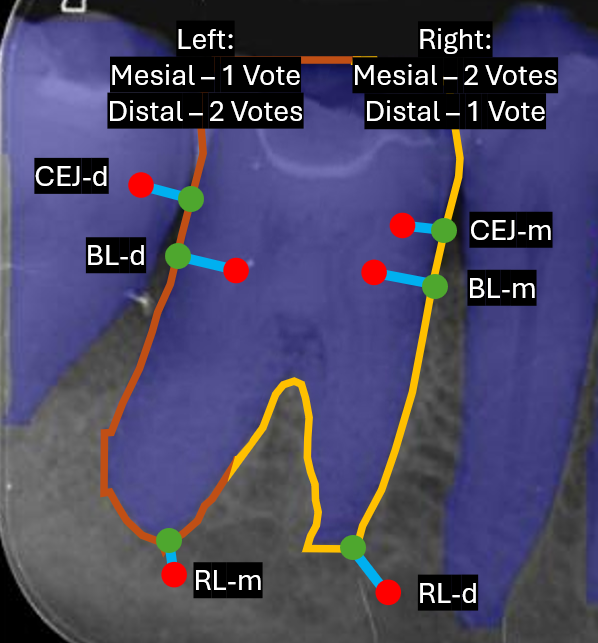}
        \caption{Mesial/Distal Determination}
        \label{subfig:postp6}
    \end{subfigure}

    \caption{Handmade example diagrams, with synthetic data, depicting each stage of the post-processing module overlaid on Image104.}
    \label{fig:postp}
\end{center}
\end{figure}

\noindent The initial stage of the post-processing module estimates the orientation of each tooth object from predicted tooth segmentation masks, in Figure \ref{subfig:postp2}. We first refine the predicted binary mask by retaining only the largest connected component, from which the boundary pixels are extracted using the Canny edge detector. Second order central moments \cite{Brodic2012} are then calculated from the edge representation of a tooth mask, where the principal axis of the tooth is then derived, giving us the orientation of the tooth mask normalised between $-90^\circ$ and $90^\circ$. This tooth orientation acts as a rotation index for subsequent tooth alignment and for evaluation purposes.

Once the orientation is derived, each tooth mask is split into left and right halves, by constructing a rotated bisector through its centroid, in Figure \ref{subfig:postp3}. This step ensures that all subsequent anatomical assignments are performed relative to the true orientation of each tooth independently, rather than the image axes.

The post-processing module then determines the mask edge pixel associations for each predicted keypoint to the anatomical boundaries of each predicted tooth. The lowest Euclidean distance for a given keypoint to each closest edge pixel for all mask halves is chosen. This provides a geometric link between anatomical landmarks and boundary structures, allowing precise localisation relative to tooth edges. This step can be seen in Figure \ref{subfig:postp4}, where the red point is a single example of a predicted keypoint, the green point is the closest edge pixel for the closest segmentation mask halve, and yellow points are the closest edge pixel for all other mask halves.

To resolve which mask best corresponds to a given group of keypoints, the average distance between visible keypoints and their associated edges is computed. The mask with the smallest mean distance is selected as the most appropriate match, effectively filtering out false positive masks.

At this stage we have many keypoint-edge associations for all predicted keypoints, where each keypoint has an association for every predicted left and right side segmentation mask in the image. The post-processing module then proceeds to match each predicted keypoint group with its most likely associated predicted mask, by choosing the mask with the lowest average distance across all visible keypoints in the group. This process filters out false postive mask predictions, by disregarding non matched masks. This process can be seen in Figure \ref{subfig:postp5}, where green points are the closest edge pixel for a given predicted red keypoint, and yellow keypoints are the edge pixels for the matched segmentation mask if that keypoint's closest edge pixel is not with the matched mask.  

Finally, the post-processing module determines which sides of the image are mesial and distal. These are determined by distance based majority voting across visible keypoints for mesial and distal related keypoints. Each keypoint is assigned to the closest side of the matched mask, and votes are accumulated by anatomical keypoint class. The side with the most votes for mesial keypoints defines the mesial half, with the distal side assigned as its inverse. The final post-processed keypoints replace raw predictions when appropriate, ensuring consistent anatomical alignment across all teeth, as shown in Figure \ref{subfig:postp6}.

Supplementary equations on these post-processing steps are in \ref{app_postp}.

\subsection{Percentage of Relative Correct Keypoints Metric}

We propose an evaluation metric, Percentage of Relative Correct Keypoints ($PRCK$), which is based on the Percentage of Correct Keypoints ($PCK$) metric \cite{yang2013computational} for pose estimation. $PCK$ measures the proportion of predicted keypoints that lie within a specified threshold distance from their corresponding ground-truth keypoints, normalised by a reference scale. A higher $PCK$ score indicates more accurate predictions, while a threshold closer to $1.0$ makes the measure more lenient. Although both $PCK$ and $PRCK$ share this principle, the key difference lies in the choice of the normalising factor $L$. Standard $PCK$ typically defines $L$ using a task-specific reference length, such as the head size or torso length in human pose estimation for each object individually, while $PRCK$ defines $L$ as the average bounding box diagonal distance across all objects in a given image, ensuring a consistent scale across objects of different sizes within an image.

In human pose estimation, using $PCK$ with body-part-specific scales is effective because keypoint distances naturally compress or expand depending on body orientation (e.g., lateral vs. anterior facing), and the metric should reflect these differences. However, in our domain, images may contain multiple teeth with substantially different sizes and anatomies, while needing to retain the same scale for accurate periodontal keypoint evaluation. Using object-specific normalisation would unfairly penalise smaller single-root teeth compared to larger multi-root teeth. Instead adopting the average bounding box diagonal as the normalising factor, $PRCK$ standardises the evaluation across all teeth in the image, grounding the metric to the relative average object size while ensuring fair comparison between structures. The $PRCK$ metric is formally defined in Equation \eqref{PRCKeq}.

\begin{equation}
\begin{split}
	\text{PRCK} = \frac{1}{n} \sum_{i=1}^{n}{\mathbbm{1}[ || y_i - \hat{y}_i ||^{2}_{2}} < d_{thresh} \cdot L] \,\,,
	\label{PRCKeq}
\end{split}
\end{equation}

where $n$ is the number of keypoints for a single class, $y_i$ is the target keypoint, $\hat{y}_i$ is the predicted keypoint, $d_{thresh}$ is a specified threshold value between $0$ and $1$, and $L$ is the normalising factor that grounds the metric to a domain. $\mathbbm{1}[\cdot]$ is an indicator function that returns $1$ if the condition is true and $0$ if the condition is false. The $d_{thresh}$ values for our evaluation is ($0.5$, $0.25$, $0.05$) and $L$ is the average diagonal distance for all tooth boxes in the image.

\section{Experiments and Results}

\subsection{Experimental Setup}

\subsubsection{Model Setup with Post-Processing}

\begin{figure}[htbp]
    \centering \includegraphics[width=1.00\linewidth]{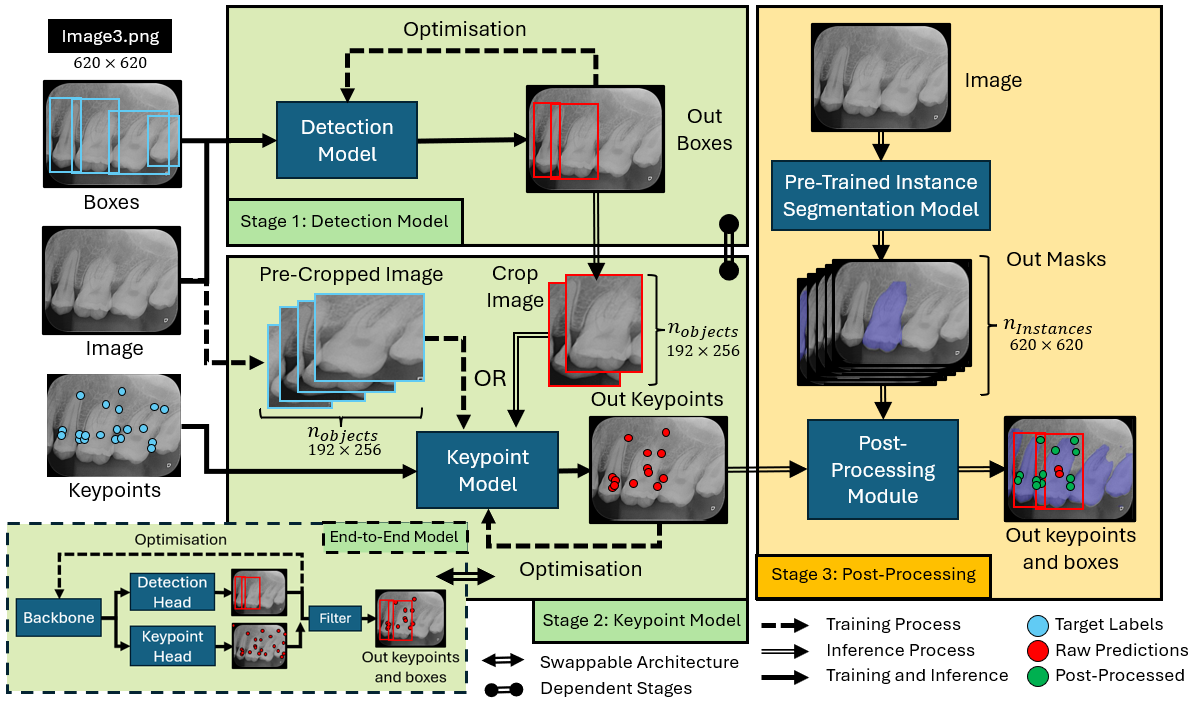}
    \caption{Training/inference loops for the two-stage top-down pipeline with post-processing, and the end-to-end YOLOv8 variant.}
    \label{fig:model_diagram}
\end{figure}

Our proposed post-processing module is model-agnostic, as long as the donor model has a matched keypoint and bounding box output per image. Figure \ref{fig:model_diagram} illustrates how our approach integrates both two-stage and end-to-end keypoint detectors. In the two-stage setting, Stage 1 is an object detector that localises tooth objects and is trained independently. Stage 2 is a single-object keypoint detector that, during training, receives pre-cropped images of the tooth objects, applies strong random augmentation, and predicts one keypoint per keypoint class for each crop. During inference, predictions from Stage 1 define the input regions for Stage 2 using detected tooth objects for the crop. Therefore, multi-object capability is enabled by the object detector. Alternatively, when using an end-to-end keypoint detector such as YOLOv8-Pose, Stages 1 and 2 are replaced by a unified architecture with object detection and keypoint heads optimised jointly. Finally, Stage 3 performs post-processing to refine the predicted keypoints by combining the raw detections with heuristic domain knowledge and outputs from a pre-trained tooth instance segmentation model.segmentation model.

\subsubsection{Model Choices and Experimental Setup}

We evaluate four donor architectures for periodontal landmark detection: an end-to-end regression model (YOLOv8-Pose) and three two-stage top-down models (DeepPose (ResNet50+RLE), HRNet, RTMPose-tiny). For the two stage models, we adopt RTMDet-tiny \cite{lyu2022rtmdet} as the object detector. All are pre-trained on COCO 2017 \cite{lin2014coco} and fine-tuned on our dataset. Input sizes are resized to $620\times620$ for stage 1 and end-to-end detectors, with $192\times256$ inputs for stage 2 keypoint models. 

From $192$ images, we hold out $17$ as a test set. The remaining $175$ images were used to train five independent models per architecture, under a $5$-fold cross-validation scheme, each with identical tuned hyperparameters for each fold. Each fold contains an alternating train/val split of $140$/$35$. We report average fold-wise validation performance on our dataset, on an external validation set \cite{altukroni2023additional} annotated using our protocol, and provide hold-out test results in \ref{app1}. Dataset splits are further explained in Table \ref{tab:splits}

\begin{table}
    \centering
    \caption{Table containing images and model per dataset splits per fold, where each fold contains a different splits from our randomised dataset. All data from test and external sets are evaluated on all 5 of the models. Exact dataset splits can be found in our dataset in Section \ref{sec:data}}
    \scriptsize
    \begin{tabular}{c|ccccc}
        \toprule
         Dataset & Fold 0 & Fold 1 & Fold 2 & Fold 3 & Fold 4\\
         \midrule
         Model & 1 & 2 & 3 & 4 & 5 \\
         \midrule
         Train (140) & (36-175) & (1-35, 71-175) & (1-70, 106-175) & (1-105, 141-175) & (1-140) \\
         Validation (35) & (1-35) & (36-70) & (71-105) & (106-140) & (141-175) \\
         Test (17) & (17) & (17) & (17) & (17) & (17) \\
         \midrule
         External & (15) & (15) & (15) & (15) & (15)\\
         \bottomrule
    \end{tabular}
    \label{tab:splits}
\end{table}

Full training and augmentation hyperparameters for each model are shown in \ref{app:train_details}.

\subsubsection{Metrics}

We assess model performance on both our dataset and an external validation set across three tasks: bounding box detection, keypoint detection, and localised disease classification. Bounding box and keypoint detection metrics are comprised of commonly used evaluation metrics and our proposed metric, while the disease classification metrics measure and evaluates the bone loss stage for the mesial and distal probe sites of each visible tooth.

\textbf{Bounding Box and Tooth Rotation Metrics}

Bounding box performance is primarily evaluated using mean Average Precision (mAP),  
$\text{mAP}(t) = \frac{1}{clss} \sum_{i=1}^{clss} \text{AP}_i(t)$, where $clss$ is the number of classes and $\text{AP}_i(t)$ the area under the precision–recall curve at IoU threshold $t$. We report mAP at $0.5$ and $[0.5:0.95]$. The former reflects detection with lenient IoU overlap, while the latter averages across thresholds to capture robustness from coarse to precise detection. We also report Precision $= \frac{TP}{TP+FP}$ and Recall $= \frac{TP}{TP+FN}$ at IoU $0.5$, where precision measures the accuracy of positive detections and Recall assesses completeness. Higher values indicate better performance.

Tooth orientation is evaluated using Normalised Mean Squared Error (NMSE), with angular differences wrapped to $(-90^\circ,90^\circ)$. This is defined as  
$\Delta = \mathrm{wrap}_{180}(\hat{\theta} - \theta) \in [-90^\circ, 90^\circ], \quad \mathrm{NMSE}_{\text{range}} = \frac{1}{N}\sum_{i=1}^N \left(\frac{\Delta_i}{90^\circ}\right)^2$,  
where $\theta_i$ is the ground truth, $\hat{\theta}_i$ the prediction, and $N$ the number of samples. Lower NMSE indicates more accurate orientation prediction.

\textbf{Keypoint Metrics}

We evaluate the keypoint performance using our proposed metric $PRCK$ in Equation \eqref{PRCKeq} at threshold values $0.5$, $0.25$, and $0.05$. Similar to the mAP metric, $PRCK$ at $0.5$ treats predicted keypoint as true positive if it is within a maximum distance of $(0.5 \cdot \text{average diagonal tooth box distance})$. Therefore, $PRCK$ at $0.5$ will indicate the general performance of the model at detecting the keypoint locations and $PRCK$ at $0.05$ will indicate the precise performance at detecting keypoint locations. The higher the value of $PRCK$, the better the performance.

\textbf{Disease Classification Metrics}

\begin{figure}[htbp]
    \begin{subfigure}{0.50\columnwidth}
        \centering \includegraphics[width=0.50\linewidth]{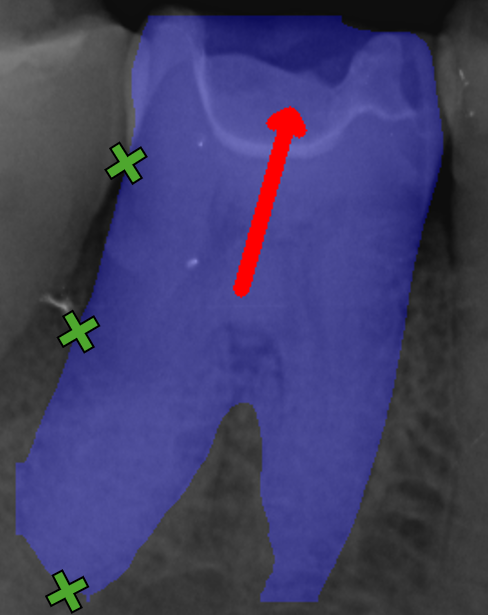}
        \caption{Keypoints}
        \label{fig:example_pbl1}
    \end{subfigure}
    \hfill
    \begin{subfigure}{0.50\columnwidth}
        \centering \includegraphics[width=0.32\linewidth]{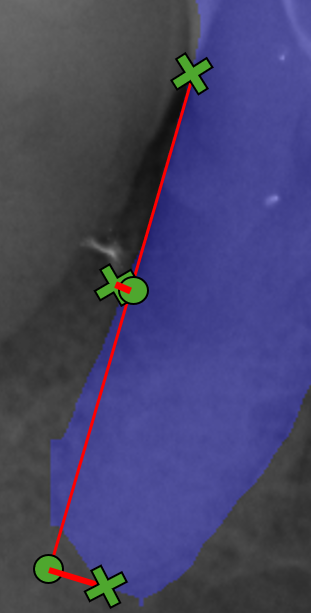}
        \caption{Projection}
        \label{fig:example_pbl2}
    \end{subfigure}

    \caption{Cropped image of Image104.png, with handmade example keypoints depicting the evaluation projection process. Keypoints are represented as green crosses, projected points as green dots, the rotation index as a red arrow and the projection from CEJ as red lines.}
    \label{fig:example_pbl}
\end{figure}

To evaluate the models' ability to detect clinically relevant stages of periodontal disease, we employ metrics that compare classification performance for conditions necessary to diagnose periodontal disease stages. To compute these metrics, we derive the Percentage of Bone Loss (PBL) using the CEJ, BL, and RL keypoints on both the distal and mesial sides of each annotated tooth. The BL and RL locations are projected along a straight line extending from the CEJ, oriented according to either the predicted rotation index or the target bounding box angle, as seen in Figure \ref{fig:example_pbl}. This gives us $\overline{BL}$ and $\overline{RL}$, which are the projected bone level and root level, respectively. We then calculate the percentage of bone loss for a specified tooth side as $\text{PBL} = \frac{||CEJ-\overline{BL}||^{2}_{2}}{||CEJ-\overline{RL}||^{2}_{2}}$, where $||CEJ-\overline{BL}||^{2}_{2}$ is the Euclidean distance of the CEJ keypoint and projected BL keypoint, and $||CEJ-\overline{RL}||^{2}_{2}$ is the Euclidean distance of the CEJ keypoint and projected RL keypoint.

PBL is calculated separately for each RL root present for a given tooth, depending on the number of roots for said tooth. To select the correct PBL value, we keep the calculation with the highest PBL value, which is done to align the percentage of bone loss calculation with the existing clinical practice of measuring PBL from the shortest root. With the localised percentage of bone loss values for both predicted and target keypoints, we can generate a multi-class confusion matrix with classes $\text{PBL}<0.15$, $0.15\leq\text{PBL}<0.33$, $0.33\leq\text{PBL}<0.66$ and $\text{PBL}\geq0.66$, as Healthy, Mild, Moderate and Severe, respectively. While clinical PBL measurements are taken from $(\text{CEJ} - 2mm)$, we cannot accurately determine the exact position of the healthy bone level due to the anonymous nature of the dataset, the unknown x-ray receptor guide dimensions, and the lack of software fixing of foreshortening/elongation artefacts.

We facilitate furcation involvement classification metrics by measuring the Euclidean distance of the each FBL keypoints to the FA keypoint, independently. If any one of the FBL keypoints has an Euclidean distance greater that $0.05$ of the average diagonal distance for all tooth bounding boxes in the image, it is considered to have furcation involvement. A binary-class confusion matrix can be calculated using the furcation involvement classification between the predicted and target data, for each multi-root teeth in the image.

Using the confusion matrix as the basis for evaluation, we report standard classification metrics precision, recall/sensitivity, and the Dice coefficient/F1 score ($\frac{2TP}{2TP + FP + FN}$). 

Accuracy and specificity are excluded from our analysis, as they are strongly influenced by the number of true negative samples and the overall class distribution, which in computer vision domains are often highly imbalanced. An over-representation of negative samples can artificially inflate accuracy and specificity by dominating the numerator through true negative values. Additionally, balancing the frequency of appearance across periodontal disease stages is not possible without artificially sampling cases from images containing a range of disease extents. For these reasons, our evaluation emphasises balance-invariant metrics.

\subsection{Results}
\label{sec:result}

\subsubsection{Object Detection and Tooth Orientation Results}

Object detection forms the first stage and of a top-down keypoint detection pipelines, directly conditioning subsequent keypoint localisation, through cropped box input images or bounding box matching. Both RTMDet and YOLOv8 demonstrate consistently strong object detection for the most prevalent morphologies, single and double roots, as seen in Table \ref{tab:box_results_quant}. For these classes, validation mAP$^{0.5}$ is above $0.92$ for both models, which reflects the models' ability to localise the general area of these teeth. This also extends beyond coarse localisation, where double root teeth for YOLOv8 achieved $mAP^{0.75}$ of $0.941$($\pm 0.025$) and $mAP^{0.5:0.95}$ of $0.755$($\pm 0.032$), and for RTMDet achieved $mAP^{0.75}$ of $0.922$($\pm 0.066$) and $mAP^{0.5:0.95}$ of $0.741$($\pm 0.052$) on the validation set, indicating accurate bounding box placement and stability across stricter IoU thresholds.

\begin{table}[!htbp]
\centering
\caption{Class-wise results for RTMDet and YOLOv8 object detection models. Results are reported as mean($\pm$standard deviation) for validation and external datasets. Standard deviation is calculated over 5-fold validation sets, and the whole external set on the individual 5-fold models.}
\scriptsize
\begin{tabular}{l l|ccc|ccc}
\toprule
\multicolumn{8}{c}{Bounding Box Evaluation}\\
\midrule
\multirow{3}{*}{Model} & \multirow{3}{*}{Class} & \multicolumn{3}{c|}{Validation} & \multicolumn{3}{c}{External} \\
\cmidrule(lr){3-5} \cmidrule(lr){6-8}
 & & mAP$^{0.5}$ & mAP$^{0.75}$ & mAP$^{0.5:0.95}$ & mAP$^{0.5}$ & mAP$^{0.75}$ & mAP$^{0.5:0.95}$ \\
\midrule
\multirow{5}{*}{RTMDet} &
 Single Root &
 \shortstack{$0.943$\\($\pm0.017$)} &
 \shortstack{$0.845$\\($\pm0.063$)} &
 \shortstack{$0.675$\\($\pm0.028$)} &
 \shortstack{$0.824$\\($\pm0.068$)} &
 \shortstack{$0.249$\\($\pm0.074$)} &
 \shortstack{$0.361$\\($\pm0.022$)} \\
 &
 Double Root &
 \shortstack{$0.956$\\($\pm0.038$)} &
 \shortstack{$0.922$\\($\pm0.066$)} &
 \shortstack{$0.741$\\($\pm0.052$)} &
 \shortstack{$0.864$\\($\pm0.015$)} &
 \shortstack{$0.222$\\($\pm0.140$)} &
 \shortstack{$0.362$\\($\pm0.033$)} \\
 &
 Triple Root &
 \shortstack{$0.833$\\($\pm0.180$)} &
 \shortstack{$0.828$\\($\pm0.189$)} &
 \shortstack{$0.644$\\($\pm0.170$)} &
 \shortstack{$0.550$\\($\pm0.173$)} &
 \shortstack{$0.000$\\($\pm0.000$)} &
 \shortstack{$0.177$\\($\pm0.083$)} \\
 &
 ARR &
 \shortstack{$0.634$\\($\pm0.060$)} &
 \shortstack{$0.317$\\($\pm0.242$)} &
 \shortstack{$0.363$\\($\pm0.121$)} &
 \shortstack{$0.228$\\($\pm0.101$)} &
 \shortstack{$0.050$\\($\pm0.105$)} &
 \shortstack{$0.088$\\($\pm0.045$)} \\
 &
 PLS &
 \shortstack{$0.160$\\($\pm0.064$)} &
 \shortstack{$0.024$\\($\pm0.026$)} &
 \shortstack{$0.059$\\($\pm0.031$)} &
 \shortstack{$0.226$\\($\pm0.083$)} &
 \shortstack{$0.030$\\($\pm0.045$)} &
 \shortstack{$0.079$\\($\pm0.046$)} \\
\midrule
\multirow{5}{*}{YOLOv8} &
 Single Root &
 \shortstack{$0.928$\\($\pm0.025$)} &
 \shortstack{$0.851$\\($\pm0.038$)} &
 \shortstack{$0.670$\\($\pm0.030$)} &
 \shortstack{$0.835$\\($\pm0.041$)} &
 \shortstack{$0.221$\\($\pm0.046$)} &
 \shortstack{$0.352$\\($\pm0.026$)} \\
 &
 Double Root &
 \shortstack{$0.966$\\($\pm0.017$)} &
 \shortstack{$0.941$\\($\pm0.025$)} &
 \shortstack{$0.755$\\($\pm0.032$)} &
 \shortstack{$0.880$\\($\pm0.115$)} &
 \shortstack{$0.235$\\($\pm0.153$)} &
 \shortstack{$0.361$\\($\pm0.073$)} \\
 &
 Triple Root &
 \shortstack{$0.839$\\($\pm0.079$)} &
 \shortstack{$0.784$\\($\pm0.107$)} &
 \shortstack{$0.622$\\($\pm0.099$)} &
 \shortstack{$0.497$\\($\pm0.445$)} &
 \shortstack{$0.000$\\($\pm0.000$)} &
 \shortstack{$0.129$\\($\pm0.116$)} \\
 &
 ARR &
 \shortstack{$0.678$\\($\pm0.026$)} &
 \shortstack{$0.302$\\($\pm0.074$)} &
 \shortstack{$0.318$\\($\pm0.031$)} &
 \shortstack{$0.580$\\($\pm0.308$)} &
 \shortstack{$0.361$\\($\pm0.279$)} &
 \shortstack{$0.330$\\($\pm0.186$)} \\
 &
 PLS &
 \shortstack{$0.164$\\($\pm0.014$)} &
 \shortstack{$0.025$\\($\pm0.023$)} &
 \shortstack{$0.060$\\($\pm0.009$)} &
 \shortstack{$0.233$\\($\pm0.049$)} &
 \shortstack{$0.028$\\($\pm0.021$)} &
 \shortstack{$0.084$\\($\pm0.028$)} \\
\midrule
\end{tabular}
\label{tab:box_results_quant}
\end{table}

The triple root class, although more anatomically complex and less frequent in the dataset, were still detected with a validation performance of $mAP^{0.5}$ at $0.839$($\pm0.079$) and $mAP^{0.75}$ at $0.784$($\pm0.107$) for YOLOv8, and $mAP^{0.5}$ at $0.833$($\pm0.180$) and $mAP^{0.75}$ at $0.828$($\pm0.189$). This still indicates a strong ability to detect triple root teeth, but with a reduced quality of detected boxes for some samples, as stipulated by the increased standard deviation. Additionally, ARR and PLS classes further indicates the difficulty of detecting under-represented classes with object detection models. For example, validation $mAP^{0.5:0.95}$ for the ARR class is $0.363$($\pm0.121$) for RTMDet and $0.318$($\pm0.031$) for YOLOv8, implying high false positive predictions with lower quality predicted bounding boxes compared to tooth boxes. RTMDet achieved a validation $mAP^{0.5}$ of $0.160$($\pm0.064$) for PLS, and YOLOv8 $mAP^{0.5}$ of $0.164$($\pm0.014$), indicating significant failure at detecting PLS conditions, despite containing more samples than triple root and ARR classes.

Despite bounding box variability, orientation prediction remains stable, with NMSE below $0.0054$ for the validation set and $0.0193$ for the external set, as seen in Table \ref{tab:orientation_results}. While all models perform consistently well, single root teeth have better validation performance below $0.0028$, compared to double (below $0.0093$) and triple (below $0.0043$) root teeth, likely due to increased elongation of single root teeth. This indicates that our post-processing method can accurately predict tooth orientation from tooth segmentation masks, despite variation in object detection performance. Although the tooth orientation methodology uses a deterministic central moments method from the same YOLOv8-Pose tooth instance segmentation model, tooth masks are discarded if not matched to a predicted tooth bounding box, hence the variability in orientation performance.

\begin{table}[!htbp]
\centering
\caption{Orientation NMSE results comparing RTMDet and YOLOv8 models across root types, for validation and external datasets. Results are reported as mean($\pm$standard deviation), calculated over 5-folds.}
\scriptsize
\begin{tabular}{l|cc|cc}
\toprule
\multicolumn{5}{c}{Orientation Evaluation} \\
\midrule
\multirow{2}{*}{Category} & \multicolumn{2}{c|}{Validation NMSE} & \multicolumn{2}{c}{External NMSE} \\
\cmidrule(lr){2-3} \cmidrule(lr){4-5}
 & RTMDet & YOLOv8 & RTMDet & YOLOv8 \\
\midrule
Single Root   & \shortstack{$0.0028$\\($\pm0.0004$)} & \shortstack{$0.0022$\\($\pm0.0007$)} & \shortstack{$0.0394$\\($\pm0.0518$)} & \shortstack{$0.0011$\\($\pm0.0001$)} \\
Double Root   & \shortstack{$0.0093$\\($\pm0.0031$)} & \shortstack{$0.0074$\\($\pm0.0025$)} & \shortstack{$0.0143$\\($\pm0.0006$)} & \shortstack{$0.0146$\\($\pm0.0005$)} \\
Triple Root   & \shortstack{$0.0041$\\($\pm0.0017$)} & \shortstack{$0.0043$\\($\pm0.0012$)} & \shortstack{$0.0043$\\($\pm0.0011$)} & \shortstack{$0.0032$\\($\pm0.0008$)} \\
Average       & \shortstack{$0.0054$\\($\pm0.0035$)} & \shortstack{$0.0046$\\($\pm0.0027$)} & \shortstack{$0.0193$\\($\pm0.0333$)} & \shortstack{$0.0063$\\($\pm0.0060$)} \\
\bottomrule
\end{tabular}
\label{tab:orientation_results}
\end{table}

\begin{figure}[!htb]
\begin{center}
    \begin{subfigure}{0.32\columnwidth}
        \centering \includegraphics[width=.94\linewidth]{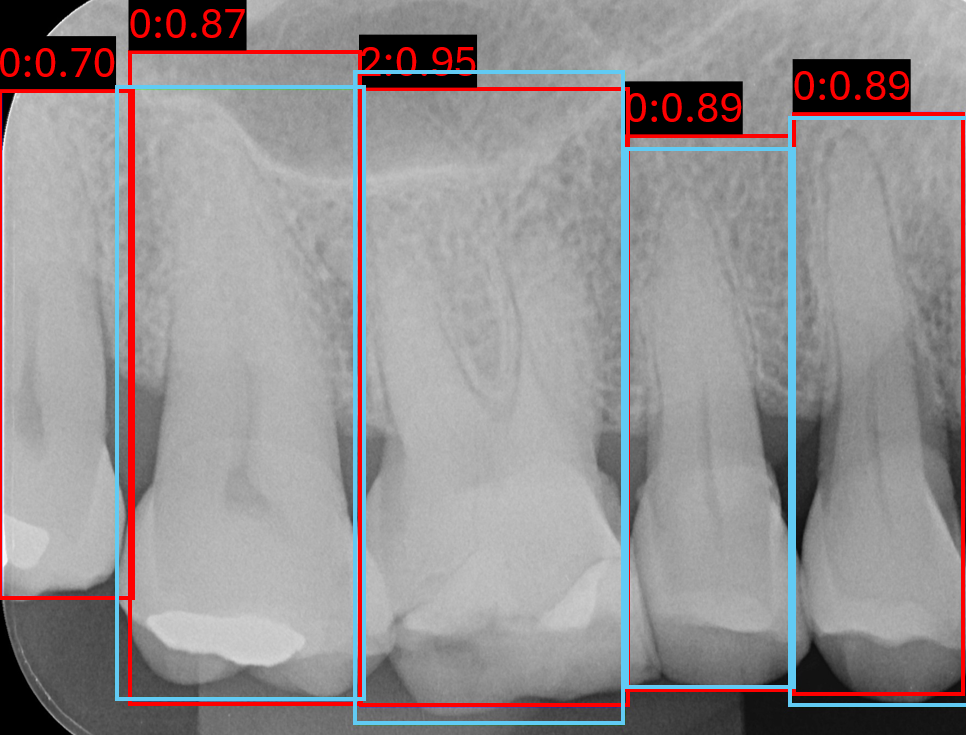}
        \caption{Image 119 YOLOv8}
        \label{subfig:box_qual_a}
    \end{subfigure}
    \begin{subfigure}{0.32\columnwidth}
        \centering \includegraphics[width=1.00\linewidth]{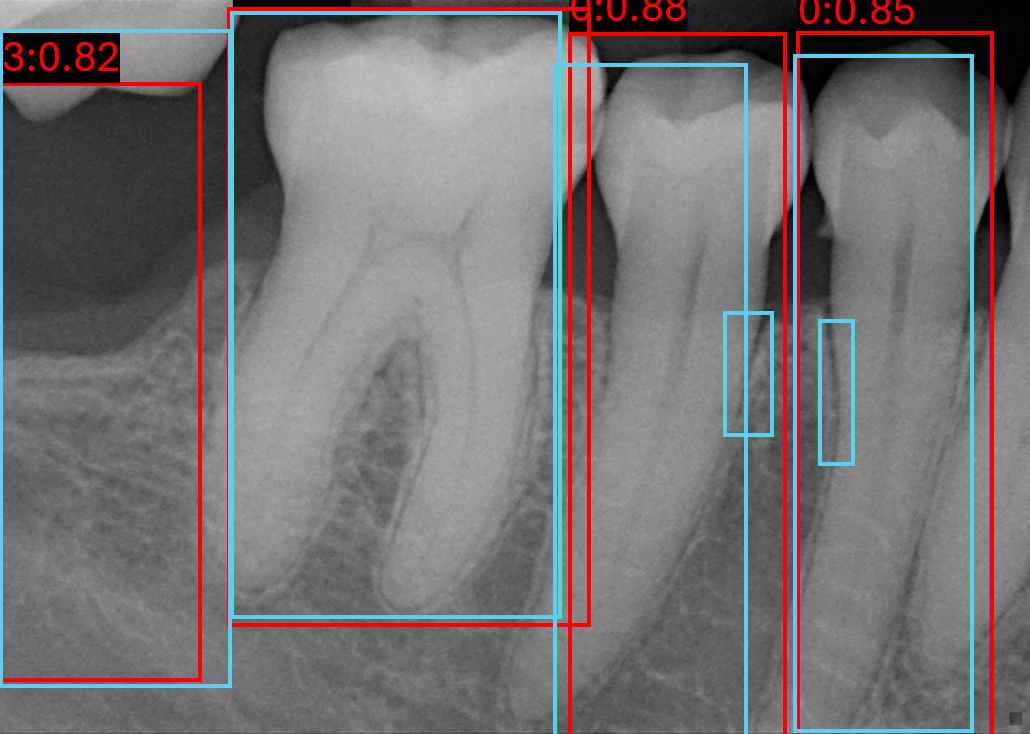}
        \caption{Image 120 YOLOv8}
        \label{subfig:box_qual_b}
    \end{subfigure}
    \begin{subfigure}{0.32\columnwidth}
        \centering \includegraphics[width=.95\linewidth]{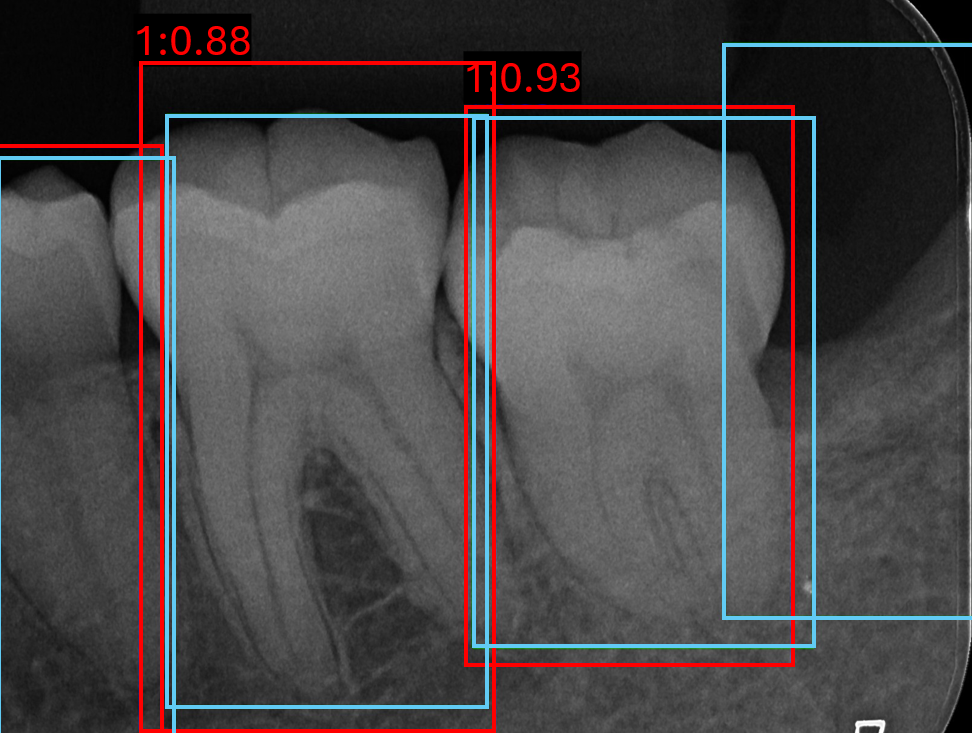}
        \caption{Image 171 YOLOv8}
        \label{subfig:box_qual_c}
    \end{subfigure}
    \newline
    \begin{subfigure}{0.32\columnwidth}
        \centering \includegraphics[width=.94\linewidth]{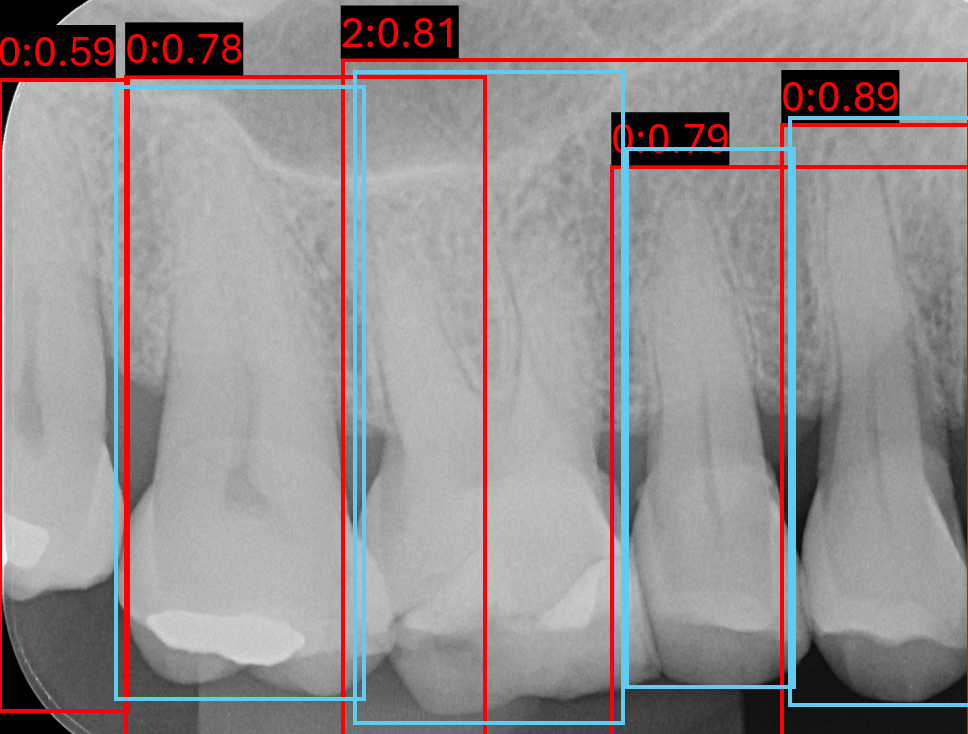}
        \caption{Image 119 RTMDet}
        \label{subfig:box_qual_d}
    \end{subfigure}
    \begin{subfigure}{0.32\columnwidth}
        \centering \includegraphics[width=1.00\linewidth]{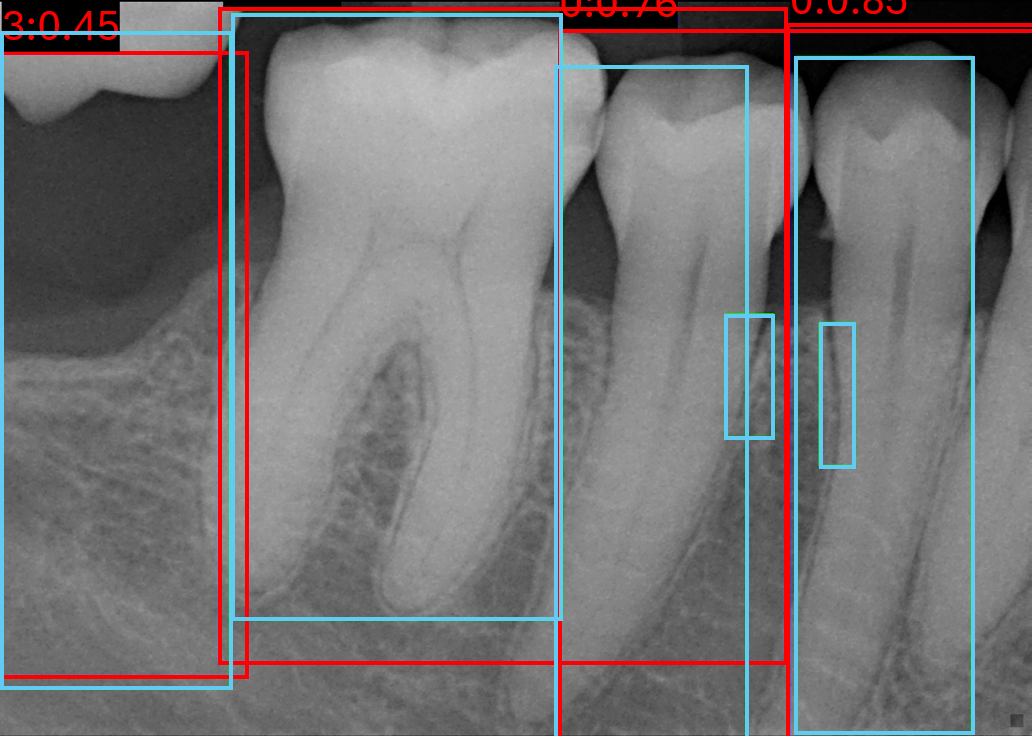}
        \caption{Image 120 RTMDet}
        \label{subfig:box_qual_e}
    \end{subfigure}
    \begin{subfigure}{0.32\columnwidth}
        \centering \includegraphics[width=.94\linewidth]{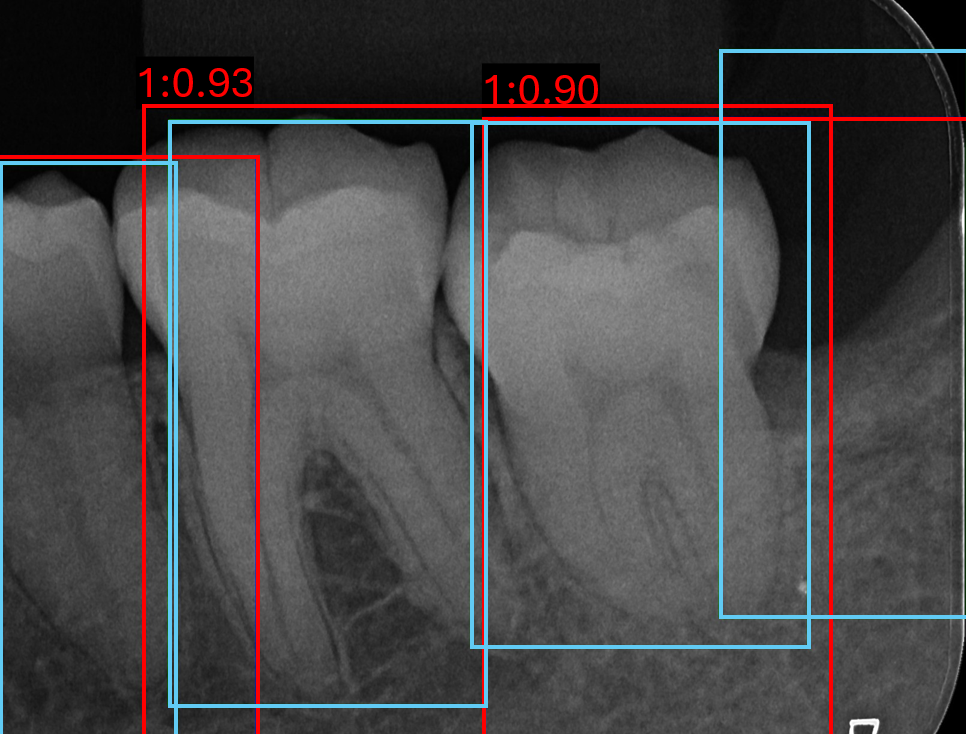}
        \caption{Image 171 RTMDet}
        \label{subfig:box_qual_f}
    \end{subfigure}

    \caption{Six validation images with overlaid bounding box results, where light blue is the target boxes and red is the predicted boxes.}
    \label{fig:box_qual}
\end{center}
\end{figure}

Figure \ref{fig:box_qual} shows qualitative examples of detection performance on the validation set. Both models generally localise tooth boundaries with high precision, although analysing RTMDet performance indicates occasionally grouping of multiple teeth within a single bounding box, while still often retaining the appropriate number of boxes per tooth in the image. In some cases, both methods detect teeth absent from the annotations, suggesting improved actual sensitivity relative to the ground truth but at the cost of reduced quantitative precision.

Both models consistently fail on PLS classes, despite its higher sample size compared to ARR and triple root classes, shown in Figure \ref{subfig:box_qual_b} and Figure \ref{subfig:box_qual_e}. This limitation is likely due to the visual similarity between healthy and widened PLS cases, indicative of a more challenging detection problem. In contrast, ARR and triple root teeth exhibit richer and more distinctive features, leading to stronger performance overall. However, false negative predictions for ARR remain evident throughout, in Figure \ref{subfig:box_qual_c} and Figure \ref{subfig:box_qual_f}.

\subsubsection{Keypoint Detection Results}

Keypoint localisation was evaluated using $PRCK$ across multiple thresholds, with and without post-processing in Table \ref{tab:prck_results}. At the coarse threshold $PRCK^{0.5}$, YOLOv8 outperforms all other models, achieving $0.912$($\pm0.026$) on the validation set and $0.900$($\pm0.029$) on the external set, reflecting strong robustness against localisation error. In contrast, performance at the strict threshold PRCK$^{0.05}$ indicates HRNet’s advantage in fine-grained precision compared to its lower generalised precision at lower thresholds, where it achieved the highest scores of $0.375$($\pm0.027$) on the validation set and $0.405$($\pm0.029$) on the external set.

\begin{table}[!htb]
\centering
\caption{Table containing $PRCK$ keypoint results for all models, with and without post-processing, for the validation and external sets, at thresholds $0.5$, $0.25$, and $0.05$. Results are reported as mean($\pm$standard deviation), where standard deviation is calculated over 5-folds.}
\scriptsize
\begin{tabular}{l|ccc|ccc}
\toprule
\multicolumn{7}{c}{Keypoint Evaluation}\\
\midrule
\multirow{3}{*}{Model} & 
\multicolumn{3}{c|}{No Post-Processing} & 
\multicolumn{3}{c}{Post-Processing} \\
\cmidrule(lr){2-4} \cmidrule(lr){5-7}
& PRCK$^{0.5}$ & PRCK$^{0.25}$ & PRCK$^{0.05}$ 
& PRCK$^{0.5}$ & PRCK$^{0.25}$ & PRCK$^{0.05}$ \\
\midrule
& \multicolumn{6}{c}{\textbf{Validation}} \\
\midrule
DeepPose   & \shortstack{$ \textbf{0.694}$\\($\pm 0.020$)}& \shortstack{$ \textbf{0.539}$\\($\pm 0.030$)}& \shortstack{$ 0.095 $\\($\pm 0.029$)}& \shortstack{$ 0.649 $\\($\pm 0.033$)}& \shortstack{$ 0.431 $\\($\pm 0.027$)}& \shortstack{$ \textbf{0.162}$\\($\pm 0.026$)}\\ 
HRNet   & \shortstack{$ \textbf{0.678}$\\($\pm 0.029$)}& \shortstack{$ \textbf{0.499}$\\($\pm 0.026$)}& \shortstack{$ \textbf{0.375}$\\($\pm 0.027$)}& \shortstack{$ 0.647 $\\($\pm 0.041$)}& \shortstack{$ 0.476 $\\($\pm 0.044$)}& \shortstack{$ 0.326 $\\($\pm 0.041$)}\\ 
RTMPose   & \shortstack{$ \textbf{0.626}$\\($\pm 0.023$)}& \shortstack{$ \textbf{0.352}$\\($\pm 0.023$)}& \shortstack{$ 0.086 $\\($\pm 0.028$)}& \shortstack{$ 0.597 $\\($\pm 0.026$)}& \shortstack{$ 0.307 $\\($\pm 0.018$)}& \shortstack{$ \textbf{0.102}$\\($\pm 0.032$)}\\ 
YOLOv8   & \shortstack{$ \textbf{0.912}$\\($\pm 0.026$)}& \shortstack{$ \textbf{0.763}$\\($\pm 0.040$)}& \shortstack{$ 0.368 $\\($\pm 0.059$)}& \shortstack{$ 0.903 $\\($\pm 0.027$)}& \shortstack{$ 0.729 $\\($\pm 0.033$)}& \shortstack{$ \textbf{0.404}$\\($\pm 0.048$)}\\ 
\midrule
& \multicolumn{6}{c}{\textbf{External}} \\
\midrule
DeepPose   & \shortstack{$ \textbf{0.865}$\\($\pm 0.019$)}& \shortstack{$ \textbf{0.683}$\\($\pm 0.050$)}& \shortstack{$ 0.091 $\\($\pm 0.014$)}& \shortstack{$ 0.805 $\\($\pm 0.034$)}& \shortstack{$ 0.579 $\\($\pm 0.027$)}& \shortstack{$ \textbf{0.194}$\\($\pm 0.033$)}\\ 
HRNet   & \shortstack{$ \textbf{0.862}$\\($\pm 0.024$)}& \shortstack{$ \textbf{0.626}$\\($\pm 0.022$)}& \shortstack{$ \textbf{0.405}$\\($\pm 0.029$)}& \shortstack{$ 0.810 $\\($\pm 0.038$)}& \shortstack{$ 0.583 $\\($\pm 0.029$)}& \shortstack{$ 0.379 $\\($\pm 0.050$)}\\ 
RTMPose   & \shortstack{$ \textbf{0.786}$\\($\pm 0.025$)}& \shortstack{$ \textbf{0.431}$\\($\pm 0.020$)}& \shortstack{$ 0.084 $\\($\pm 0.017$)}& \shortstack{$ 0.742 $\\($\pm 0.031$)}& \shortstack{$ 0.405 $\\($\pm 0.035$)}& \shortstack{$ \textbf{0.107}$\\($\pm 0.020$)}\\ 
YOLOv8   & \shortstack{$ \textbf{0.900}$\\($\pm 0.029$)}& \shortstack{$ \textbf{0.702}$\\($\pm 0.039$)}& \shortstack{$ 0.309 $\\($\pm 0.083$)}& \shortstack{$ 0.894 $\\($\pm 0.026$)}& \shortstack{$ 0.667 $\\($\pm 0.069$)}& \shortstack{$ \textbf{0.356}$\\($\pm 0.086$)}\\ 
\bottomrule
\end{tabular}
\label{tab:prck_results}
\end{table}

Post-processing consistently improved strict-threshold $PRCK^{0.05}$ performance across models. For example, YOLOv8 increased from $0.368$($\pm0.059$) to $0.404$($\pm0.048$) on the validation set and from $0.309$($\pm0.083$) to $0.356$($\pm0.086$) on the external set. However, these gains in fine localisation were often accompanied by reductions at broader thresholds, $PRCK^{0.25}$ and $PRCK^{0.5}$.

\begin{figure}[!htb]
\begin{center}
    \begin{subfigure}{0.24\columnwidth}
        \centering \includegraphics[width=1.0\linewidth]{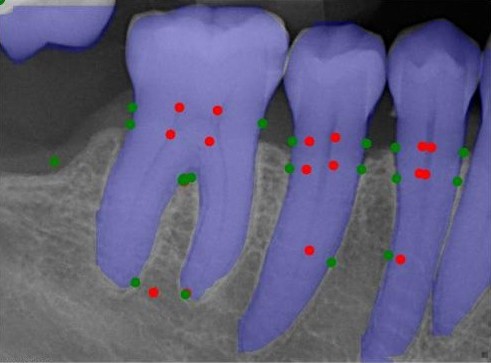}
        \caption{Image 120 DeepPose}
        \label{subfig:kpt_qual_a}
    \end{subfigure}
    \begin{subfigure}{0.24\columnwidth}
        \centering \includegraphics[width=1.00\linewidth]{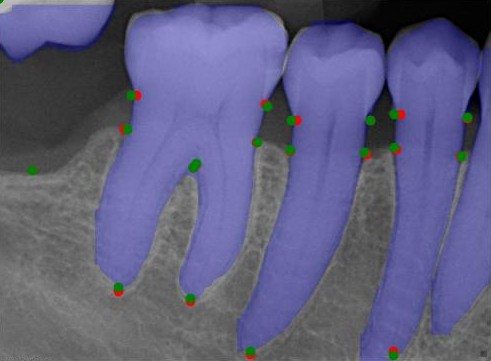}
        \caption{Image 120 HRNet}
        \label{subfig:kpt_qual_b}
    \end{subfigure}
    \begin{subfigure}{0.24\columnwidth}
        \centering \includegraphics[width=1.0\linewidth]{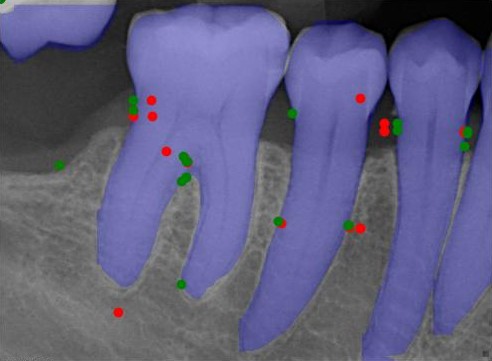}
        \caption{Image 120 RTMPose}
        \label{subfig:kpt_qual_c}
    \end{subfigure}
    \begin{subfigure}{0.24\columnwidth}
        \centering \includegraphics[width=1.0\linewidth]{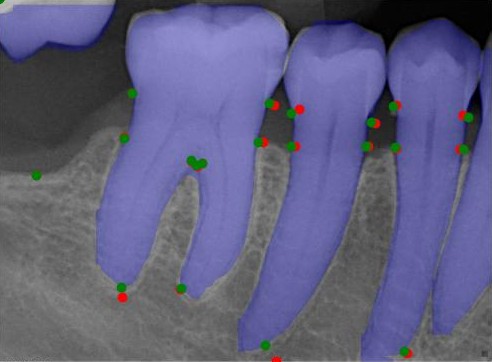}
        \caption{Image 120 YOLOv8}
        \label{subfig:kpt_qual_d}
    \end{subfigure}
    \newline
    \begin{subfigure}{0.24\columnwidth}
        \centering \includegraphics[width=1.0\linewidth]{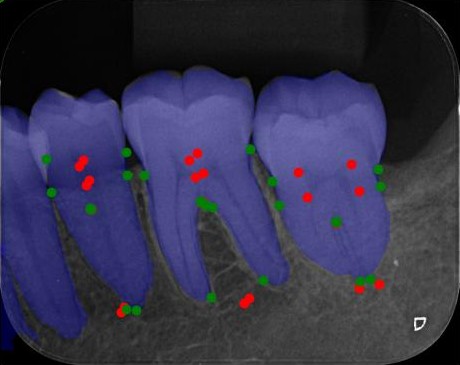}
        \caption{Image 171 DeepPose}
        \label{subfig:kpt_qual_e}
    \end{subfigure}
    \begin{subfigure}{0.24\columnwidth}
        \centering \includegraphics[width=1.00\linewidth]{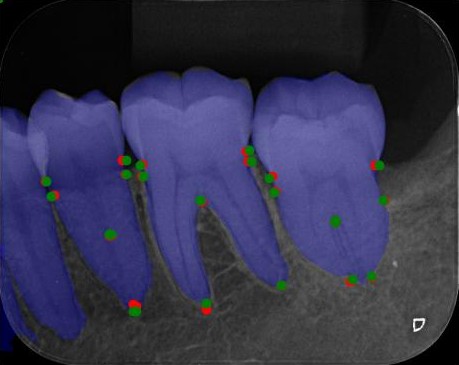}
        \caption{Image 171 HRNet}
        \label{subfig:kpt_qual_f}
    \end{subfigure}
    \begin{subfigure}{0.24\columnwidth}
        \centering \includegraphics[width=1.0\linewidth]{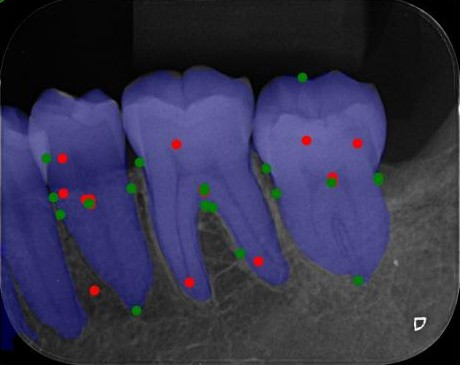}
        \caption{Image 171 RTMPose}
        \label{subfig:kpt_qual_g}
    \end{subfigure}
    \begin{subfigure}{0.24\columnwidth}
        \centering \includegraphics[width=1.0\linewidth]{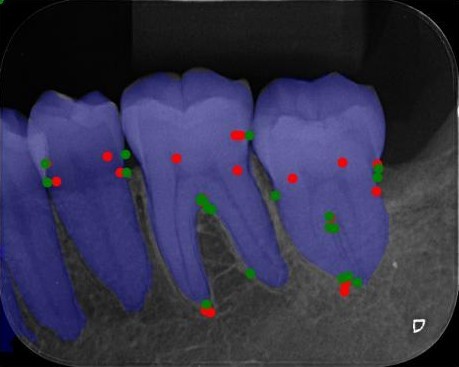}
        \caption{Image 171 YOLOv8}
        \label{subfig:kpt_qual_h}
    \end{subfigure}

    \caption{Six validation images with overlay keypoint results, where red points are the raw keypoint predictions and green points are the post-processed keypoints.}
    \label{fig:kpt_qual_qual}
\end{center}
\end{figure}

Analysing the validation set, in Figure \ref{fig:kpt_qual_qual}, qualitatively shows that raw predictions rarely coincide with anatomically correct locations, often being detected within the tooth interior or entirely outside its boundary. Post-processing substantially improves localisation in most cases, shifting keypoints towards plausible mesial and distal edges, as shown in Figure \ref{subfig:kpt_qual_a} and \ref{subfig:kpt_qual_d}. However, this refinement is heavily dependent on the quality of the raw detections. When predictions are excessively noisy, post-processing can amplify errors, relocating keypoints to implausible locations such as the crown or furcation apex, as seen in Figure \ref{subfig:kpt_qual_g} and Figure \ref{subfig:kpt_qual_h}. This further explains the observed quantitative increase for $prck^{0.05}$, but declination at more lenient thresholds, since small adjustments improve low-tolerance metrics, yet fail to increase high-tolerance thresholds.

\begin{figure}[htbp]
    \centering \includegraphics[width=.85\linewidth]{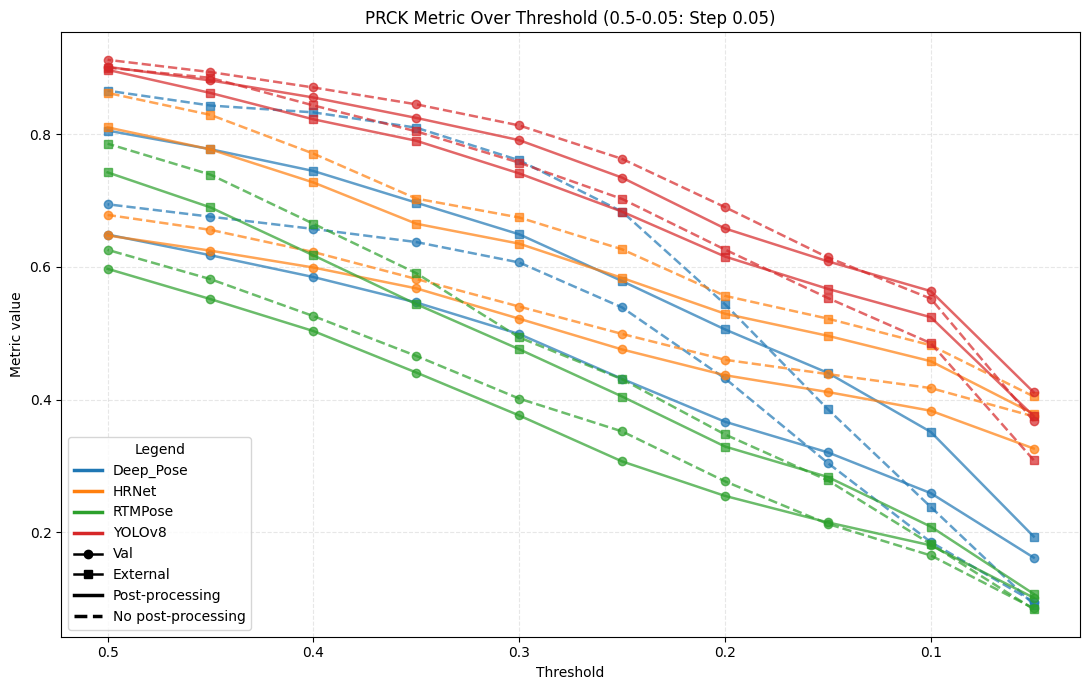}
    \caption{$PRCK$ metrics for the validation and external set, with and without post-processing, for all keypoint detection models, over all thresholds from $0.5$ to $0.05$ with a step of $0.05$.}
    \label{fig:prck_plot}
\end{figure}

Further analysis of post-processing $PRCK$ metrics at a range of thresholds in Figure \ref{fig:prck_plot}, all models except HRNet show slightly reduced $0.5$-$0.2$ threshold performance for post-processed keypoints compared to no post-processing. However, post-processed and non-post-processed performance inverts between a threshold of $0.2$-$0.1$.

\subsubsection{Localised Disease Classification Results}

The PBL localised classification results in Table \ref{tab:pbl_results} shows that HRNet and YOLOv8 consistently achieve the strongest overall performance, with the highest Dice scores across both mesial ($0.508$ validation) and distal ($0.464$ validation) evaluations in both datasets. DeepPose performs moderately well but remains below HRNet and YOLOv8, while RTMPose shows comparatively weaker results. Post-processing generally leads to performance degradation for all models on the validation set. However, RTMPose and DeepPose benefits from post-processing on the external set, particularly in precision and Dice. These suggests that, while post-processing tends to negatively affect models that already perform strongly, it can enhance weaker models such as RTMPose by stabilising predictions and reducing variability.

\begin{table}[!htb]
\centering
\caption{Mesial and distal side Percent Bone Loss classification results with and without post-processing, over all models, for the validation and external datasets. Metrics are reported as mean($\pm$standard deviation), where standard deviation is calculated over 5-folds.}
\scriptsize
\begin{tabular}{l|ccc|ccc}
\toprule
\multicolumn{7}{c}{Percentage of Bone Loss Evaluation}\\
\midrule
\multirow{3}{*}{Model} & 
\multicolumn{3}{c|}{No Post-Processing} & 
\multicolumn{3}{c}{Post-Processing} \\
\cmidrule(lr){2-4} \cmidrule(lr){5-7}
& Precision & Recall & Dice (F1)
& Precision & Recall & Dice (F1) \\

\midrule
& \multicolumn{6}{c}{\textbf{Validation - Mesial}} \\
\midrule
DeepPose & \shortstack{$\textbf{0.416}$\\($\pm0.080$)} & \shortstack{$\textbf{0.431}$\\($\pm0.091$)} & \shortstack{$\textbf{0.395}$\\($\pm0.095$)} & \shortstack{$0.345$\\($\pm0.064$)} & \shortstack{$0.362$\\($\pm0.075$)} & \shortstack{$0.338$\\($\pm0.070$)} \\
HRNet    & \shortstack{$\textbf{0.517}$\\($\pm0.059$)} & \shortstack{$\textbf{0.522}$\\($\pm0.060$)} & \shortstack{$\textbf{0.508}$\\($\pm0.059$)} & \shortstack{$0.495$\\($\pm0.037$)} & \shortstack{$0.501$\\($\pm0.031$)} & \shortstack{$0.489$\\($\pm0.033$)} \\
RTMPose  & \shortstack{$\textbf{0.293}$\\($\pm0.112$)} & \shortstack{$\textbf{0.128}$\\($\pm0.114$)} & \shortstack{$\textbf{0.124}$\\($\pm0.048$)} & \shortstack{$0.249$\\($\pm0.109$)} & \shortstack{$0.121$\\($\pm0.094$)} & \shortstack{$0.106$\\($\pm0.018$)} \\
YOLOv8   & \shortstack{$\textbf{0.477}$\\($\pm0.035$)} & \shortstack{$\textbf{0.416}$\\($\pm0.019$)} & \shortstack{$\textbf{0.425}$\\($\pm0.023$)} & \shortstack{$0.444$\\($\pm0.017$)} & \shortstack{$0.412$\\($\pm0.017$)} & \shortstack{$0.419$\\($\pm0.020$)} \\

\midrule
& \multicolumn{6}{c}{\textbf{Validation - Distal}} \\
\midrule
DeepPose & \shortstack{$\textbf{0.382}$\\($\pm0.049$)} & \shortstack{$\textbf{0.383}$\\($\pm0.077$)} & \shortstack{$\textbf{0.365}$\\($\pm0.064$)} & \shortstack{$0.322$\\($\pm0.057$)} & \shortstack{$0.325$\\($\pm0.045$)} & \shortstack{$0.311$\\($\pm0.051$)} \\
HRNet    & \shortstack{$\textbf{0.476}$\\($\pm0.044$)} & \shortstack{$\textbf{0.465}$\\($\pm0.034$)} & \shortstack{$\textbf{0.464}$\\($\pm0.040$)} & \shortstack{$0.447$\\($\pm0.046$)} & \shortstack{$0.440$\\($\pm0.056$)} & \shortstack{$0.441$\\($\pm0.051$)} \\
RTMPose  & \shortstack{$\textbf{0.277}$\\($\pm0.052$)} & \shortstack{$\textbf{0.243}$\\($\pm0.138$)} & \shortstack{$\textbf{0.185}$\\($\pm0.044$)} & \shortstack{$0.207$\\($\pm0.068$)} & \shortstack{$0.161$\\($\pm0.060$)} & \shortstack{$0.152$\\($\pm0.038$)} \\
YOLOv8   & \shortstack{$\textbf{0.451}$\\($\pm0.038$)} & \shortstack{$\textbf{0.417}$\\($\pm0.024$)} & \shortstack{$\textbf{0.423}$\\($\pm0.022$)} & \shortstack{$0.429$\\($\pm0.067$)} & \shortstack{$0.399$\\($\pm0.043$)} & \shortstack{$0.404$\\($\pm0.046$)} \\

\midrule
& \multicolumn{6}{c}{\textbf{External - Mesial}} \\
\midrule
DeepPose & \shortstack{$0.355$\\($\pm0.095$)} & \shortstack{$0.395$\\($\pm0.115$)} & \shortstack{$0.343$\\($\pm0.111$)} & \shortstack{$\textbf{0.394}$\\($\pm0.055$)} & \shortstack{$\textbf{0.452}$\\($\pm0.069$)} & \shortstack{$\textbf{0.395}$\\($\pm0.051$)} \\
HRNet    & \shortstack{$\textbf{0.484}$\\($\pm0.086$)} & \shortstack{$\textbf{0.574}$\\($\pm0.146$)} & \shortstack{$\textbf{0.510}$\\($\pm0.103$)} & \shortstack{$0.447$\\($\pm0.042$)} & \shortstack{$0.514$\\($\pm0.075$)} & \shortstack{$0.464$\\($\pm0.046$)} \\
RTMPose  & \shortstack{$0.283$\\($\pm0.092$)} & \shortstack{$\textbf{0.217}$\\($\pm0.107$)} & \shortstack{$0.113$\\($\pm0.067$)} & \shortstack{$\textbf{0.329}$\\($\pm0.125$)} & \shortstack{$0.190$\\($\pm0.142$)} & \shortstack{$\textbf{0.144}$\\($\pm0.044$)} \\
YOLOv8   & \shortstack{$\textbf{0.511}$\\($\pm0.038$)} & \shortstack{$\textbf{0.496}$\\($\pm0.045$)} & \shortstack{$\textbf{0.493}$\\($\pm0.024$)} & \shortstack{$0.410$\\($\pm0.060$)} & \shortstack{$0.464$\\($\pm0.100$)} & \shortstack{$0.427$\\($\pm0.072$)} \\

\midrule
& \multicolumn{6}{c}{\textbf{External - Distal}} \\
\midrule
DeepPose & \shortstack{$\textbf{0.337}$\\($\pm0.120$)} & \shortstack{$\textbf{0.350}$\\($\pm0.117$)} & \shortstack{$\textbf{0.319}$\\($\pm0.123$)} & \shortstack{$0.330$\\($\pm0.083$)} & \shortstack{$0.327$\\($\pm0.094$)} & \shortstack{$0.310$\\($\pm0.093$)} \\
HRNet    & \shortstack{$\textbf{0.459}$\\($\pm0.056$)} & \shortstack{$\textbf{0.519}$\\($\pm0.048$)} & \shortstack{$\textbf{0.473}$\\($\pm0.052$)} & \shortstack{$0.452$\\($\pm0.080$)} & \shortstack{$0.497$\\($\pm0.094$)} & \shortstack{$0.464$\\($\pm0.086$)} \\
RTMPose  & \shortstack{$0.241$\\($\pm0.096$)} & \shortstack{$0.247$\\($\pm0.064$)} & \shortstack{$0.142$\\($\pm0.060$)} & \shortstack{$\textbf{0.338}$\\($\pm0.098$)} & \shortstack{$\textbf{0.328}$\\($\pm0.125$)} & \shortstack{$\textbf{0.215}$\\($\pm0.050$)} \\
YOLOv8   & \shortstack{$\textbf{0.509}$\\($\pm0.108$)} & \shortstack{$\textbf{0.407}$\\($\pm0.034$)} & \shortstack{$\textbf{0.412}$\\($\pm0.049$)} & \shortstack{$0.432$\\($\pm0.101$)} & \shortstack{$0.385$\\($\pm0.066$)} & \shortstack{$0.389$\\($\pm0.067$)} \\
\bottomrule
\end{tabular}
\label{tab:pbl_results}
\end{table}

The furcation involvement results in Table \ref{tab:furcation_results} show a clear discrepancy between the classification of healthy cases and those with furcation involvement. For healthy sites, all models achieve strong performance across both validation and external datasets, with all metric values consistently above $0.87$. Although the detection of diseased furcation involvement shows very low performance for all models. DeepPose completely fails to identify diseased cases, yielding zero scores across all metrics, while HRNet, RTMPose, and YOLOv8 achieve only marginal improvements, with highly unstable precision and recall, as a result of extremely low instance counts for diseased furcation areas.

\begin{table}[!htb]
\centering
\caption{Healthy and diseased furcation involvement results over all models, for the validation and external datasets. Metrics are reported as mean($\pm$standard deviation), where standard deviation is calculated over 5-folds.}
\scriptsize
\begin{tabular}{l|ccc|ccc}
\toprule
\multicolumn{7}{c}{Furcation Involvement Evaluation}\\
\midrule
\multirow{3}{*}{Model} & 
\multicolumn{3}{c|}{Validation} & 
\multicolumn{3}{c}{External} \\
\cmidrule(lr){2-4} \cmidrule(lr){5-7}
& Precision & Recall & Dice (F1)
& Precision & Recall & Dice (F1) \\
\midrule
& \multicolumn{6}{c}{\textbf{Healthy}} \\
\midrule
DeepPose   & \shortstack{$0.885$\\($\pm0.063$)} & \shortstack{$1.000$\\($\pm0.000$)} & \shortstack{$0.938$\\($\pm0.035$)} & \shortstack{$0.872$\\($\pm0.004$)} & \shortstack{$1.000$\\($\pm0.000$)} & \shortstack{$0.931$\\($\pm0.002$)} \\
HRNet   & \shortstack{$0.891$\\($\pm0.056$)} & \shortstack{$1.000$\\($\pm0.000$)} & \shortstack{$0.942$\\($\pm0.030$)} & \shortstack{$0.883$\\($\pm0.025$)} & \shortstack{$1.000$\\($\pm0.000$)} & \shortstack{$0.938$\\($\pm0.014$)} \\
RTMPose   & \shortstack{$0.898$\\($\pm0.071$)} & \shortstack{$0.951$\\($\pm0.048$)} & \shortstack{$0.922$\\($\pm0.043$)} & \shortstack{$0.880$\\($\pm0.028$)} & \shortstack{$0.969$\\($\pm0.038$)} & \shortstack{$0.922$\\($\pm0.029$)} \\
YOLOv8   & \shortstack{$0.900$\\($\pm0.049$)} & \shortstack{$0.855$\\($\pm0.056$)} & \shortstack{$0.875$\\($\pm0.032$)} & \shortstack{$0.879$\\($\pm0.032$)} & \shortstack{$0.888$\\($\pm0.085$)} & \shortstack{$0.881$\\($\pm0.045$)} \\
\midrule
& \multicolumn{6}{c}{\textbf{Furcation Involvement}} \\
\midrule
DeepPose   & \shortstack{$0.000$\\($\pm0.000$)} & \shortstack{$0.000$\\($\pm0.000$)} & \shortstack{$0.000$\\($\pm0.000$)} & \shortstack{$0.000$\\($\pm0.000$)} & \shortstack{$0.000$\\($\pm0.000$)} & \shortstack{$0.000$\\($\pm0.000$)} \\
HRNet   & \shortstack{$0.200$\\($\pm0.400$)} & \shortstack{$0.040$\\($\pm0.080$)} & \shortstack{$0.067$\\($\pm0.133$)} & \shortstack{$0.200$\\($\pm0.400$)} & \shortstack{$0.100$\\($\pm0.200$)} & \shortstack{$0.133$\\($\pm0.267$)} \\
RTMPose   & \shortstack{$0.200$\\($\pm0.400$)} & \shortstack{$0.133$\\($\pm0.267$)} & \shortstack{$0.160$\\($\pm0.320$)} & \shortstack{$0.200$\\($\pm0.400$)} & \shortstack{$0.100$\\($\pm0.200$)} & \shortstack{$0.133$\\($\pm0.267$)} \\
YOLOv8   & \shortstack{$0.142$\\($\pm0.133$)} & \shortstack{$0.200$\\($\pm0.194$)} & \shortstack{$0.162$\\($\pm0.149$)} & \shortstack{$0.100$\\($\pm0.200$)} & \shortstack{$0.100$\\($\pm0.200$)} & \shortstack{$0.100$\\($\pm0.200$)} \\
\bottomrule
\end{tabular}
\label{tab:furcation_results}
\end{table}

\newpage

\section{Discussion and Conclusion}

In this study we propose a novel dataset annotation methodology, for the detection of periodontal disease related keypoints and defects, with the goal of supplying fast and accurate information for a clinician to make an informed diagnosis. We also propose a keypoint detection metric for dental imaging domains, Percentage of Relative Correct Keypoints, that is based on the Percentage of Detected Joints metric, but normalises the metric to the average tooth size of present teeth in the image. Lastly, we propose a post-processing technique that moves certain keypoint predictions to the edge of the related tooth.

Periodontal disease treatment is based on an accurate classification of the disease to achieve an adequate diagnosis, prognosis, and treatment plan that minimise possible human errors. The diagnosis of periodontal disease is made through clinical and radiographic analysis, which can be subjective in some cases. The development of artificial intelligence tools could assist clinicians in optimising care for each patient. Advances in artificial intelligence are increasing due to the increased digitisation of radiography (and healthcare in general), the development of novel algorithms and neural network architectures, and the increasing computational power available. The advantage of our proposed annotation methodology is that periodontists would not have to manually calculate bone loss percentages for each tooth, a very time-consuming and sometimes inaccurate process.

The complexity of making a complete diagnosis and prediction of periodontal disease using only two-dimensional periapical radiographs. For a more accurate diagnosis and prediction, it is necessary to thoroughly review the clinical and radiographic data together, including the patient's history, clinical depth on probing, clinical attachment loss, bleeding on probing, and tooth mobility. Oral radiologists are highly qualified professionals, so any tool that can maximise the quality and efficiency of this process is of great interest. Artificial intelligence can also improve image quality by enhancing image reconstruction and filtering equipment, such as volumetric computed tomography, \cite{kida2018cone} potentially improving spatial resolution. Therefore, an algorithm using only periapical radiographic images does not provide sufficient evidence, though it can still serve as a reference for diagnosis.

The data annotation methodology of using keypoints to detect periodontal bone loss, allows for stage agnostic detection and increases annotation counts. Detecting periodontal disease stages only using a stage dependent method such as object detection would likely result in poor performance with disease stage classes with small instance counts, as seen with PLS detection. Additionally, the method of detection provides a clear and easily understood method for clinicians, as it automates a process that is already performed manually. The method also allows for easy identification and correction of false predictions in real world scenarios, as incorrect keypoint locations can be easily moved to the correct location, while the increased obfuscated decision making of solely object detection methods will likely cause confusion when false detections are predicted.

Our results demonstrate strong overall object detection and keypoint performance across all evaluated models. Post-processing offers marginal improvements for strict thresholds ($PRCK^{0.05}$), though it slightly reduces performance under more lenient metrics ($PRCK^{0.25}$ and $PRCK^{0.5}$). However, qualitatively post-processing substantially enhances clinical readability of keypoint locations, even when quantitative metrics appear unchanged or occasionally degraded. External validation further supports the clinical relevance of the annotation methodology, with most models performing comparably or better than the validation set. 

Among the evaluated donor models, YOLOv8-Pose is the best performing model due to its increased object detection performance and end-to-end architecture, while HRNet achieves the most accurate qualitative localisation of keypoints when only considering its positive samples. However, performance for PLS and furcation involvement is notably weaker for all models, likely reflecting insufficient representation of these cases within the dataset. 

Post-processing failures are particularly evident when keypoints are processed towards the crown or furcation apex. Addressing this in accordance with our methodology may require further segmenting teeth into crown and root masks, enabling easier context-aware filtering of these regions. Additionally, an approach that embeds anatomical priors directly into the training process, such as incorporating tooth boundary information or enforcing topological consistency, could produce more robust predictions and reduce reliance on heuristic post-processing. While our post-processing provides consistent qualitative improvements when raw predictions are reasonable, its dependence on predicted keypoint quality limits its robustness. The clinical significance of this method could also be further explored in clinical settings, even though scalability is proven for most tasks due to similar external and validation performance. Methodological or data improvements for PLS and furcation detection tasks are essential for further research in this area, because of their consistently low performance. Therefore, future work should primarily focus on integrating explicit anatomical constraints into model design and increasing the performance or number of under-represented conditions in the dataset.

\section*{Data Availability}
\label{sec:data}

\noindent The annotated dataset for this paper can be found on Zenodo \cite{guerrero2025periodontal}, here: 

\noindent\url{https://zenodo.org/records/17272200} 

\section*{Code Availability}
\label{sec:code}
\noindent The code used to train and evaluate our method can be found here: 

\noindent\url{https://github.com/Banksylel/Bone-Loss-Keypoint-Detection-Code}

\section*{CRediT}
\noindent \textbf{Ryan Banks}: data processing, data curation, data analysis, methodology, code, experimentation, writing – original draft, and writing – review \& editing. \textbf{Vishal Thengane}: data analysis, methodology, and writing – original draft. \textbf{María Eugenia Guerrero}: conceptualisation, dataset collection, dataset annotation, writing – original draft, and writing – review \& editing. \textbf{Nelly Maria García-Madueño}: dataset collection, dataset annotation, writing – original draft, and writing – review \& editing. \textbf{Yunpeng Li}: supervision and writing – review \& editing. \textbf{Hongying Tang}: supervision and writing – review \& editing. \textbf{Akhilanand Chaurasia}: writing – review \& editing.

\section*{Declaration of Competing Interest}
\noindent The authors declare that they have no known competing financial interests or personal relationships that could have appeared to influence the work reported in this paper.

\section*{Funding}
\noindent The dataset used in this paper was originally collected by Universidad Nacional Mayor de San Marcos under Grant A21051201. Subsequent annotation of the data was conducted by the authors of this manuscript.

\section*{Ethics Statement}

\noindent All procedures were performed in compliance with relevant laws and institutional guidelines. The collection of anonymised patient radiographs were conducted with regard to patient informed consent and data protection laws. Ethical review was carried out as part of the grant application by Universidad Nacional Mayor de San Marcos under Grant A21051201.

\section*{Use of Artificial Intelligence}
\noindent No generative artificial intelligence was used in the creation of the manuscript, figures, or artwork.

\section*{Rights Retention}
\noindent For the purpose of open access, the author has applied a Creative Commons attribution license (CC BY) to any Author Accepted Manuscript version arising from this submission.

\appendix

\section{Auxiliary Segmentation Setup and Performance}
\label{app_seg_results}

\noindent Information on the auxiliary instance segmentation model setup and auxiliary validation performance can be found in \href{supplementary1.pdf}{Supplementary Material 1 (PDF)}.

\section{Post-Processing Equations}
\label{app_postp}

\noindent A low-level explanation with equations for the post-processing module can be found in \href{supplementary2.pdf}{Supplementary Material 2 (PDF)}.

\section{Test Metrics}
\label{app1}

\noindent Evaluation of each donor model on the hold-out test set can be found in \href{supplementary3.pdf}{Supplementary Material 3 (PDF)}.

\section{Training and Augmentation Details}
\label{app:train_details}

\noindent Training and augmentation hyperparameters can be found in \href{supplementary4.pdf}{Supplementary Material 4 (PDF)}.













\end{document}